\documentclass[preprint,12pt]{elsarticle}

\usepackage{amssymb}
\usepackage{stackrel}%
\usepackage{framed}%
\usepackage{graphicx}%
\usepackage{multirow}%
\usepackage{amsmath, amssymb, amsfonts}%
\usepackage{amsthm}%
\usepackage{mathrsfs}%
\usepackage[title]{appendix}%
\usepackage{xcolor}%
\usepackage{textcomp}%
\usepackage{manyfoot}%
\usepackage{booktabs}%
\usepackage{bbding}
\usepackage{algorithm}%
\usepackage{algorithmicx}%
\usepackage{algpseudocode}%
\usepackage{listings}%
\usepackage{tabularx}%
\usepackage{caption}
\usepackage{bm}
\usepackage{float}
\usepackage{url}
\usepackage[section]{placeins}
\usepackage{threeparttable}

\theoremstyle{thmstyleone}%
\newtheorem{theorem}{Theorem}
\theoremstyle{thmstyletwo}%
\newtheorem{lemma}{Lemma}%

\theoremstyle{thmstylethree}%
\newtheorem{definition}{Definition}%

\journal{~}
\makeatletter

\newcommand{\Rmnum}[1]{\expandafter\@slowromancap\romannumeral #1@}
\makeatother

\allowdisplaybreaks[4]

\begin{document}

\begin{frontmatter}



\title{Privacy-Preserving Traceable Functional Encryption for Inner Product}


\author[label1]{Muyao Qiu} 
\author[label1]{Jinguang Han}
\affiliation[label1]{organization={School of Cyber Science and Engineering, Southeast University},
            city={Nanjing},
            postcode={210096}, 
            state={Jiangsu},
            country={China}}

\begin{abstract}
Functional encryption introduces a new paradigm that decryption only reveals the function value of encrypted data.  In order to curb key leakage issues and trace users in FE-IP, a traceable functional encryption for inner product (TFE-IP) scheme has been proposed. However, the privacy protection of user's identities has not been considered in the existing TFE-IP schemes. In order to balance privacy and accountability, we propose the concept of privacy-preserving traceable functional encryption for inner product (PPTFE-IP)  and give a concrete construction which offers the features:  (1) To prevent key sharing, both a user's identity and a vector are bound together in the key;  (2) The key generation center (KGC) and a user execute a two-party secure computing protocol to generate a key without the former knowing anything about the latter's identity;  (3) Each user can ensure the integrity and correctness of his/her key through verification;  (4)  The inner product of the two vectors embedded in a ciphertext and in his/her key can be calculated by an authorized user;  (5) Only the tracer can trace the identity embedded in a key. We formally reduce the security of the proposed PPTFE-IP to well-known complexity assumptions, and conduct an implementation to evaluate its efficiency. The novelty of our scheme is to protect users' privacy and provide traceability if required.
\end{abstract}


\begin{highlights}
\item A secret key is derived anonymously to each user by the key generation center without releasing anything about user's identity. 
\item The tracer can trace the key holder's identity if required. 
\item Definitions, security models and security proof of the proposed PPTFE-IP are formally treated. 
\end{highlights}


\end{frontmatter}



\section{Introduction}
\label{sec1}
The increasing advancements in big data and cloud computing has led to a greater emphasis on data security.  Traditional public-key cryptography only allows for an all-or-nothing decryption approach,  meaning that the decrypted data can either reveal all of the original information or nothing at all.  Functional encryption (FE) introduces a novel paradigm in public-key cryptography that decryption only reveals the function value of encrypted data~\cite{bsw:fe11}.

An instance of functional encryption is functional encryption for inner product (FE-IP),  which is only permitted to calculate the inner product of two vectors related to a user's key and a ciphertext without revealing anything else, enhancing privacy in data processing. FE-IP has numerous practical applications, such as machine learning~\cite{msh:fe19ml,rpb:fe19pml,nl23}, federated learning~\cite{xbz:fl21}, data marketing~\cite{cww:datamarketing20,kpc:datamarketing22}, Internet of Things (IoT)~\cite{sgr:iot21}, etc. 

In existing FE-IP schemes~\cite{abdp:feip15,als:feip16}, a central authority (CA) must be fully trusted to generate secret keys according to vectors for users.  Hence, the CA must be fully honest; otherwise, it can impersonate any user to decrypt ciphertexts and may release users' personal information. Han et al.~\cite{hch:ppdfeip23} presented a privacy-preserving FE-IP scheme in which the CA and a user execute a two-party secure computing protocol to generate a key without the former knowing anything about the latter's identity. Although this approach provides full anonymity in the key distribution process, it does not address the traceability problem. Therefore,  if a user shares his/her secret key with others, he/she cannot be identified.
To monitor malicious key leakage while protecting user privacy,  we propose the first privacy-preserving traceable functional encryption for inner product (PPTFE-IP) scheme. In PPTFE-IP, a tracer is introduced to identify the identity of a key holder when he/she shares his/her key with others. However, the CA knows nothing about a user's identity.

\subsection{Related Work}\label{subsec2}

In this section, we review schemes related to our scheme. 

\subsubsection{Functional Encryption}\label{subsec3}

Functional encryption (FE)~\cite{bsw:fe11} is a novel paradigm in public-key cryptography where decryption only reveals the function value of  encrypted data without revealing anything else. The concept of FE was firstly proposed by Boneh et al.~\cite{bsw:fe11},  who formalized the definition and security model of this encryption technique. 

Abdalla et al.~\cite{abdp:feip15} introduced the concept of FE-IP,  whereby the decryption process only outputs the inner product of two vectors related to a ciphertext and a user's secret key.  Abdalla et al.~\cite{abdp:feip15} presented a simple FE-IP scheme with selective security.  Building on this work,  Abdalla et al.~\cite{abdp:feip16} constructed a novel generic scheme which is secure against adaptive adversaries. Agrawal et al.~\cite{als:feip16} introduced fully secure FE-IP schemes based on the same assumptions as~\cite{abdp:feip15} and gave a simple method to achieve the bounded collusion FE for all circuits. 
Classical FE-IP schemes have a central authority (CA), deriving secret keys for users. To simplify the key management in classical FE-IP, Song et al.~\cite{hfeip21} proposed a hierarchical identity-based FE-IP, where identity has a hierarchical structure and supports delegating secret key. Han et al.~\cite{dfeip23} presented a delegatable FE-IP (DFE-IP) scheme where the delegatee is allowed to represent the delegator to decrypt ciphertexts in a specified time period.

In above FE-IP schemes, CA must be trusted fully because it keeps all the secret keys of users. This is known as the key escrow problem~\cite{keyescrow10}.  Some FE-IP schemes, such as  decentralized FE-IP~\cite{DFEIP19}, decentralized multi-client FE-IP~\cite{dmcfe18} and dynamic decentralized FE-IP~\cite{ddfeip}, were proposed mainly to reduce trust on the CA and solve the key escrow problem.  Abdalla et al.~\cite{DFEIP19} presented a decentralized FE-IP to transform any scheme with key-derivation into a decentralized key-derivation scheme. A new primitive decentralized multi-client FE-IP was proposed by Chotard et al.~\cite{dmcfe18} where clients generate ciphertexts non-interactively, and secret keys are generated in a decentralized way. J{\'e}r{\'e}my et al.~\cite{ddfeip} proposed the first dynamic decentralized FE-IP, which allows participants to join dynamically during the lifetime of a system. In these schemes, multiple authorities or clients issue keys to users, instead of one CA. 

To reduce trust on the CA and protect privacy, Han et al.~\cite{hch:ppdfeip23} introduced a decentralized privacy-preserving FE-IP scheme, in which a user and multiple authorities collaborate to generate a user's secret key, and the authorities do not know the identity associated with the user's key. 


\subsubsection{Traitor Tracing}\label{subsec3}
Traitor tracing schemes is able to trace the identity of a key holder who leaked the secret key~\cite{TT}.  There are two types of traitor tracing,  white-box tracing and black-box tracing. In white box tracing~\cite{lcw:tabe12,TABE20,TABE21}, the malicious users are traced given a well-formed secret key.  In black box tracing~\cite{bbtabe13,bbtabe14,tabe20bb}, the malicious users are traced using a black box (including an unknown key and algorithm)  which can decrypt ciphertexts.

Schemes~\cite{bbtabe13,bbtabe14} introduced black-box traceable ciphertext-policy attribute-based encryption (CP-ABE) to trace the malicious user; however, these schemes were impractical because the composite-order group is required. In order to overcome the problem, Xu et al.~\cite{tabe20bb} presented a CP-ABE scheme based on the prime-order group supporting black-box traceability.

A user's identity is embedded in his/her secret key in white-box tracing schemes~\cite{lcw:tabe12,TABE20}.  In~\cite{lcw:tabe12}, the tracer generates part of keys for users and records some auxiliary information for trace. Han et al.~\cite{TABE20} used a similar method as~\cite{lcw:tabe12} but a binary tree was applied to lower the cost of storing users' identities and corresponding information. In~\cite{TABE21}, if the secret key is leake or abused, any entity can trace the identity embedded in the key, this is called public traceability. However, restricting tracing to only the tracer is essential to ensure user privacy.  

To curb key leakage issues and trace users in FE-IP,~\cite{dpp:tfeip20} defined a novel primitive called traceable functional encryption for inner product (TFE-IP)  and proposed a concrete black-box tracing scheme for FE-IP.  Following~\cite{dpp:tfeip20}, Luo et al.~\cite{lawh:tfeip22} introduced the first efficient traceable FE-IP scheme supporting public, black-box tracing and revocation, which achieved adaptive security (A-IND-CPA) under standard assumptions.  Luo et al.~\cite{lawh:tfeip22} also proposed the first generic TFE-IP schemes that achieved adaptive security. Dutta et al.~\cite{dpsm:feip22} introduced fully collusion resistant TFE-IP schemes for the first time which were public, black-box traceable, and gave generic constructions of TFE-IP based on both pairing and lattices. The ciphertext size of their pairing-based schemes is linear with $\sqrt{n}$, while the size is linear with $n$ in lattice-based schemes, where $n$ represents the number of system users. Branco et al.~\cite{TRFEIP24} introduced a new traceable FE-IP model based on registered FE where users generate their secret-public key pairs and register their public keys. They introduced registered traitor tracing which is resistant against unbounded collusion of malicious users, and proposed registered traceable FE schemes for quadratic functions and inner product functions respectively.

However, a traceable FE-IP scheme with private traceability has not been considered. Also, users' privacy and anonymity may be violated by traceability. To balance the relationship between privacy and traceability, we propose a PPTFE-IP scheme. 

The comparison of different features between our scheme and existing schemes is illustrated in Table~\ref{tab2}.  Our scheme can recover a user's identity through a well-formed secret key,  while other schemes can trace a user's identity via decryption boxes; therefore,  our scheme is white-box traceable,  while other schemes are black-box traceable.  Among existing schemes,  schemes~\cite{lawh:tfeip22,dpsm:feip22,lawy:feip24,TRFEIP24} are publicly traceable,  which undermines user privacy.~\cite{dpp:tfeip20} provides similar trace functionality with our scheme, namely,  only the tracer can trace the identity embedded in a key,  so~\cite{dpp:tfeip20} and our scheme are private tracing; on the contrary,~\cite{lawh:tfeip22,dpsm:feip22,lawy:feip24,TRFEIP24} are public tracing. The novelty of our scheme is to protect users' privacy and provide traceability if required.

\begin{table}[h]
\caption{Comparison with Existing Work}\label{tab2}
\begin{tabular*}{\textwidth}{@{\extracolsep\fill}lccc}
 \toprule%
Scheme&Traceability&Public/Private Traceability&Privacy-Preservation \\
\midrule
~\cite{dpp:tfeip20}& Black box &Private & -- \\
~\cite{lawh:tfeip22}& Black box &Public & --\\
~\cite{dpsm:feip22}& Black box &Public & --\\
~\cite{TRFEIP24}& Black box &Public & --\\
~\cite{lawy:feip24}& Black box &Public & --\\
\bf {Ours} &\bf{ White box} &\bf{Private} &\bf{\Checkmark}\\
\bottomrule
\end{tabular*}
\end{table}

\subsection{Our Contributions}\label{subsec2}
Our PPTFE-IP scheme balances the relationship between traceability and anonymity in FE-IP. Specifically, the PPTFE-IP scheme offers some interesting features:
\begin{itemize}
\item[1]For the sake of tracing and preventing key sharing,  both a user's identity and a vector are bound together in the key. 
\item[2]For the purpose of protecting user privacy, a user and KGC jointly execute a two-party secure computing protocol to generate a secret key. KGC is unknown about the identity related to the key,  but the user can get the secret key and verify its correctness. 
\item[3] The inner product  of the two vectors embedded in a ciphertext and in his/her key can be calculated by an authorized user, without revealing other information. 
\item[4]If required, the tracer can trace the key holder's identity.  
\end{itemize}
The contributions of this paper are shown below:
\begin{itemize}
\item[1]we formalize the definitions and security models of PPTFE-IP. 
\item[2]A concrete TFE-IP scheme is developed utilizing the asymmetric pairing. 
\item[3]A PPKeyGen algorithm is proposed for TFE-IP. 
\item[4]An implementation is conducted to evaluate the efficiency of PPTFE-IP. 
\item[5]We formally reduce the security of the proposed PPTFE-IP to well-known complexity assumptions.

\end{itemize}
\subsection{Techniques and Challenges}\label{subsec2}
When constructing our PPTFE-IP scheme, We encounter the following challenges:
\begin{itemize}
\item[1]In order to implement tracing and prevent key sharing,  it is necessary to embed a user's identity in his/her key. Therefore, to protect privacy, it is challenging to generate a key for a user without knowing his/her identity. 
\item[2]The KGC and a user collaborate to generate a secret key, hence it is challenging to enable the tracer to trace the identity of the user.
\item[3]Different users may jointly generate new secret keys via collusion, and it is challenging to prevent user collusion attacks.
\item[4]In our PPTFE-IP scheme, both privacy and traceability are considered. However, it is extremely challenging to  balance the relationship between privacy and traceability.
\end{itemize}
To address these challenges, we employ the technologies as follows:
\begin{itemize}
\item[1]To protect user's identity embedded in the key, the two-party secure computing technique is applied to enable the KGC to derive a key to an anonymous user without knowing his/her identity. During key generation, the user's identity  and the generated secret key are unknown to KGC.
\item[2]To enable the tracer to trace the identity of a key holder, when generating a secret key, user's identity is encrypted by the user utilizing the public key of the tracer and a zero-knowledge proof about the encryption is provided.
\item[3]To prevent the combination of secret keys, all elements in a key are bound together by a random number. Therefore, even two users collude, they cannot combine their secret keys.
\item[4]The KGC and a user collaborate to generate a secret key, and other users and the KGC cannot learn the embedded identity from the key. Hence, the user's privacy is protected. Meanwhile, the tracer traces the identity embedded in the key when necessary.
\end{itemize}

\subsection{Organization}\label{subsec2}
We offer an overview of the preliminaries as well as the formal definition and security models of PPTFE-IP in Section~\ref{sec2}.  Section~\ref{sec3} presents the concrete construction of our PPTFE-IP scheme.  Section~\ref{sec4} offers a detailed security proof for PPTFE-IP. We conduct a comparative analysis of our scheme with existing TFE-IP schemes, implement and evaluate our scheme in Section~\ref{sec5}.  At last,  we conclude this paper and outline future work in Section~\ref{sec6}.

\section{Preliminaries}\label{sec2}
\begin{figure}[htbp]%
\centering
\includegraphics[width=0.9\textwidth]{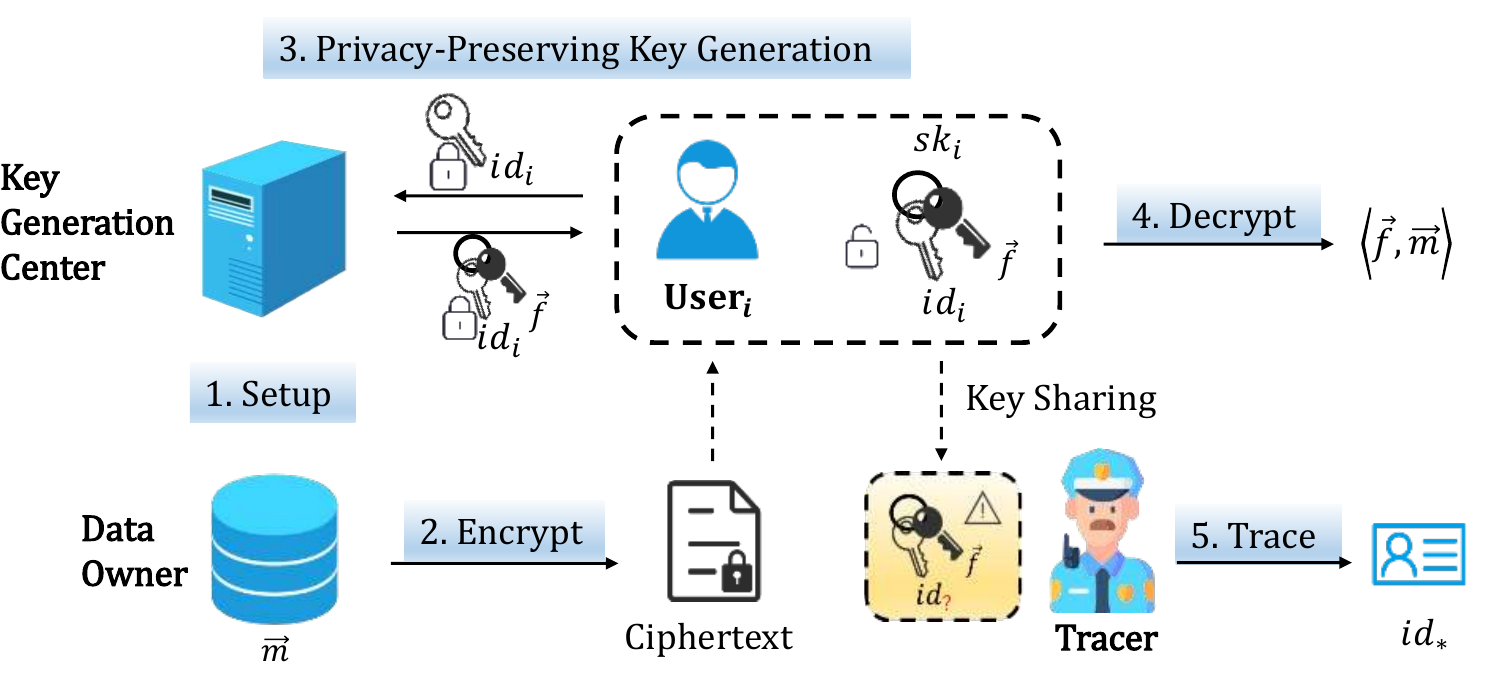}
\caption{The Framework of Our PPTFE-IP Scheme}\label{fig1}
\end{figure}

In this section, we present preliminaries along with formal definitions and security models utilized in this paper. The symbols employed in this paper are outlined in Table~\ref{tabNS}.
\begin{table}[t]
\caption{Notation Summary}\label{tabNS}
\scriptsize
\begin{tabular*}{\textwidth}{@{\extracolsep\fill}llll}
 \toprule%
Notation&Description&Notation&Description \\
\midrule
FE& Functional encryption &DL & Discrete logarithm \\
FE-IP&FE for inner product &DDH &Decisional Diffie-Hellman \\
TFE-IP& Traceable FE-IP &q-SDH & Q-strong Diffie-Hellman  \\
PPTFE-IP& Privacy-preserving TFE-IP &$1^\lambda$ & A security parameter \\
$\vec{x}$& A vector & $\langle\vec{x},\vec{y}\rangle$ & The inner product of $\vec{x}$ and $\vec{y}$\\
KGC& Key generation center &$\mathcal{BG}$ & A bilinear group generation algorithm\\
$\epsilon(\lambda)$& A negligible function in $\lambda$ &PP & Public parameters\\
$\perp$&Empty&$\zeta$&Failure\\
$com$ & Commitment &$decom$ &Decommitment\\
$\theta$ &User's identity &$\mathcal{A}$ &A probabilistically polynomial time\\ 
$\mathcal{GG}$&A group generation algorithm& &adversary\\
$\mathcal{C}$  &A challenger &$\mathcal{S}$  &A simulator\\
PPKeyGen &Privacy-preserving key generation algorithm & & \\
\bottomrule
\end{tabular*}
\end{table}
Figure~\ref{fig1} depicts the framework of our PPTFE-IP scheme.  There are three entities involved in our scheme: user,  KGC and tracer.  Each user and KGC jointly generates a secret key without exposing his/her identity.  If a malicious user discloses his/her secret key,  the tracer can trace his/her identity from the shared key. 
\subsection{Bilinear Groups and Complexity Assumptions}

\begin{definition}[Prime Order Bilinear Groups]Let $\mathbb{G}$ be a (multiplicative) cyclic group generated by $g \in \mathbb{G}$ with prime order $p$ and $\mathbb{G}_T$ be a (multiplicative) cyclic group of prime order $p$. If the following conditions are satisfied, a map $e:\mathbb{G}\times\mathbb{G}\rightarrow\mathbb{G}_T$ is a bilinear pairing:
\begin{itemize}
\item[1]Bilinearity.  $\forall a, b \in \mathbb{Z}_p,  g_1, g_2\in \mathbb{G}, e (g_1^a, g_2^b) =e (g_1^b, g_2^a) =e (g_1, g_2) ^{a\cdot b}$;
\item[2]Non-Generation.  Let 1 be the identity element in $\mathbb{G}_T. \ {\forall}g_1, g_2\in \mathbb{G},$ $ e (g_1, g_2) \neq 1$;
\item[3]Computability. ${\forall}g_1, g_2\in \mathbb{G},\ e (g_1, g_2) $ can be calculated efficiently. 
\end{itemize}
\end{definition}
\begin{definition}[Decisional Diffie-Hellman (DDH) Assumption~\cite{DH}] $\mathcal{GG}(1^\lambda)$\\
$ \rightarrow(\mathbb{G},p,g) $, where $\lambda$ is security parameter, $\mathbb{G}$ is a group generated by $g \in \mathbb{G}$ with prime order $p$.  Given $\alpha, \beta, \gamma$ randomly chosen from $\mathbb{Z}_p$, DDH assumption holds on $ (p,\mathbb{G}) $  
if the tuples $\left (g^\alpha,g^\beta,g^{\alpha\beta}\right) $ and $\left (g^\alpha,g^\beta,g^{\gamma}\right) $ are computationally indistinguishable by all adversaries $\mathcal{A}$ with a negligible advantage $\epsilon (\lambda) $, in other words,\\
\begin{center}
$Adv_{\mathcal{A}}^{DDH}=\Big|\Pr\left[\mathcal{A} (g^\alpha,g^\beta,g^{\alpha\beta}) =1\right]-\Pr\left[\mathcal{A} (g^\alpha,g^\beta,g^{\gamma}) =1\right]\Big|\leq\epsilon (\lambda) $.
\end{center}
\end{definition}
\begin{definition}[Discrete Logarithm  (DL)  Assumption~\cite{dl93}]$ \mathcal{GG}(1^\lambda) \rightarrow (\mathbb{G},p,$
$g) $, where $\lambda$ is security parameter, $\mathbb{G}$ is a group generated by $g \in \mathbb{G}$ with prime order $p$.
Given $g,y$, if for all adversaries $\mathcal{A}$ the DL assumption is satisfied on the group $(p,\mathbb{G})$, $\mathcal{A}$ have negligible advantage $\epsilon (\lambda) $ in computing $x\in\mathbb{Z}_p$ from $y=g^x$, namely\\
\begin{center}
$Adv_{\mathcal{A}}^{DL}=\Big|\Pr\left[y=g^x|\mathcal{A}\left (g,y\right) \rightarrow x) \right]\Big| \textless\epsilon (\lambda) $.
\end{center}
\end{definition}
\begin{definition}[q-Strong Diffie-Hellman  (q-SDH)  Assumption~\cite{bb04}]$ \mathcal{GG}(1^\lambda)$ \\$\rightarrow (\mathbb{G}_1,\mathbb{G}_2,p,g_1,g_2) $,\ where $\lambda$ is security parameter, $\mathbb{G}_1,\mathbb{G}_2$ are two cyclic groups of prime order $p$ respectively generated by $g_1$ and $g_2$.
Given $x$ randomly chosen from $\mathbb{Z}_p$, q-SDH assumption holds on $ (p, \mathbb{G}_1, \mathbb{G}_2) $ if given a  (q+2) -tuple $\left (g_1,g_2,g_2^x.g_2^{ (x^2) },\cdots,g_2^{ (x^q) }\right) $  
as input, output a pair $\left (c, g_1^{\frac{1}{x+c}}\right), c\in \mathbb{Z}_p$. An adversary $\mathcal{A}$ has negligible advantage $\epsilon (\lambda) $ in solving q-SDH, namely\\
\begin{center}
$Adv_{\mathcal{A}}^{q-SDH}=\Bigg|\Pr\left[\mathcal{A}\left (g_1,g_2,g_2^x,g_2^{ (x^2) },\cdots,g_2^{ (x^q) }\right) =\left (c, g_1^{\frac{1}{x+c}}\right) \right]\Bigg|\geq\epsilon (\lambda) $.
\end{center}
\end{definition}

\subsection{Formal Definition}\label{fd}
In a traceable functional encryption scheme,  the tracer can trace the identity of the corresponding key owner from the secret key. We follow the definition introduced in~\cite{dpp:tfeip20} and~\cite{lcw:tabe12} to define our TFE-IP. Firstly, the definition and security model of our TFE-IP scheme are formalized, then the formal definition and security model of our PPTFE-IP scheme are presented. 

\begin{definition}[TFE-IP]  A TFE-IP scheme is formally defined by the five algorithms as follows:
\begin{itemize}
\item[1]{\bf Setup}$\left (1^\lambda\right) \rightarrow (msk, Tk, PP) $. The Setup algorithm takes a security parameter $1^\lambda$ as input and outputs the master secret key $msk$, trace key $Tk$ and public parameters $PP$. 
\item[2]{\bf Encrypt}$ (PP, \vec{m}) \rightarrow Ct$. The Encryption algorithm inputs public parameters and a vector $\vec{m}$, produces the ciphertext $Ct$ as output. 
\item[3]{\bf KeyGen}$\left (msk, \vec{f}, id\right) \rightarrow sk_{\vec{f}, id}$. The KeyGen algorithm takes master secret key, a vector $\vec{f}$ and a user's identity $id$ as input, outputs $sk_{\vec{f}, id}$ as user's secret key for function $\vec{f}$. 
\item[4]{\bf Decrypt}$\left (Ct, sk_{\vec{f}, id}, id\right) \rightarrow \langle \vec{f}, \vec{m}\rangle\ or\perp$. The Decryption algorithm takes ciphertext, user's secret key and user's id as input,  decrypts the inner product value. If the decryption failed, the algorithm outputs $\perp$. 
\item[5]{\bf Trace}$\left (sk_{\vec{f}, id}, Tk\right) \rightarrow id\ or\perp$. The Tracing algorithm takes user's secret key $sk_{\vec{f}, id}$ and tracer's trace key as input, outputs the user's $id$ or failure $\perp$. 
\end{itemize}
\end{definition}

\subsection{Security Model}
\subsubsection{s-IND-CPA Security}
To define the security, we apply the selective indistinguishability against chosen plaintext attacks  (s-IND-CPA)  model. The game below executed between a challenger $\mathcal{C}$ and an adversary $\mathcal{A}$ define this s-IND-CPA  model.
\begin{itemize}
\item {\bf Initialization. } Two vectors $\vec{m_0}, \vec{m_1}$ with the same length are submitted by the adversary $\mathcal{A}$. 
\item {\bf Setup. } The challenger $\mathcal{C}$ runs $\left (1^\lambda\right) \rightarrow (msk, PP) $ and returns $PP$ to $\mathcal{A}$ and creates a set $V$ which is initially empty.  
\item {\bf Phase-I  (KeyGen Query) . } $\mathcal{A}$ submits an identity $id$ and a vector $\vec{f}$ limiting $\langle\vec{f}, \vec{m_0}\rangle=\langle\vec{f}, \vec{m_1}\rangle$.  $\mathcal{C}$ executes $KeyGen (msk,\vec{f}, id) $ \\$\rightarrow sk_{\vec{f}, id}$,  and returns $sk_{\vec{f}, id}$.  $\mathcal{C}$ updates $V\leftarrow V\cap\{\vec{f}, id\}$.  The adversary $\mathcal{A}$ queries multiple times. 
\item {\bf Challenge. } $\mathcal{C}$ randomly selects a $\delta \in \{0, 1\}$.  $\mathcal{C}$ executes $Enc (PP, m_\delta) \rightarrow Ct$, and returns $Ct$ to $\mathcal{A}$. 
\item {\bf Phase-II. } {\bf Phase-I} is repeated. 
\item {\bf Output. } $\mathcal{A}$ guesses $\delta^\prime$ about $\delta$.  In the case that $\delta^\prime=\delta$,  $\mathcal{A}$ wins the game. 
\end{itemize}
\begin{definition}[s-IND-CPA Security] A traceable functional encryption for inner product scheme is secure in the s-IND-CPA model if any  adversary $\mathcal{A}$ has a negligible advantage $\epsilon(\lambda)$ in winning the above game,  namely
\begin{center}
$Adv_{\mathcal{A}}=\left|\Pr[\delta^\prime=\delta]-\frac{1}{2}\right|<\epsilon (\lambda) $. 
\end{center}
\end{definition}

\subsubsection{Traceability}
For traceability, we follow the security model proposed in~\cite{lcw:tabe12,ndc:tabe15,ncd:tabe16}. This security model is described by the game as below which is executed by a challenger $\mathcal{C}$ and an adversary $\mathcal{A}$.
\begin{itemize}
\item {\bf Setup. } $\mathcal{C}$ executes ${\bf Setup}\left (1^\lambda\right) $, sending public parameters $PP$ to $\mathcal{A}$. 
\item {\bf Key Query. } $\mathcal{A}$ submits $ (id, \vec{f}) $ in which $id$ represents identity and  $\vec{f}$ represents function.  $\mathcal{C}$ returns the secret key matching the queried pair to $\mathcal{A}$. $\mathcal{A}$ can query multiple times.
\item {\bf Trace Query. } $\mathcal{A}$ sends a key $ (sk_i) _{i\in[r]}=  ({K_1}_i, {K_2}_i,$ ${K_3}_i, {K_4}_i, {K_5}_i) $ to $\mathcal{C}$.  $\mathcal{C}$ runs {\bf Trace}$\left (sk_i, Tk\right) $ to recover the identity and sends it to $\mathcal{A}$. $\mathcal{A}$ can query multiple times.
\item {\bf Key Forgery. } $\mathcal{A}$ forges and produces $sk_*$ as output.  Assume the identity embedded in $sk_*$ is $id_*$. $\mathcal{A}$ wins the game if $Trace\left (sk_*, tsk, PP\right) \neq id_*$ or\ $Trace\left (sk_*, tsk, PP\right)\notin\{\bot, id_1, id_2,$ $\cdots, id_q\}$. 
\end{itemize}
\begin{definition}[Traceability] A TFE-IP scheme is fully traceable if any adversary $\mathcal{A}$ wins the above game with a negligible advantage $\epsilon(\lambda)$,  namely
\begin{center}
$Adv_{\mathcal{A}}=\Pr[\left\{Trace\left (sk^*, tsk, PP\right) \notin\{\bot, id_1, id_2,\cdots, id_q\}\right\}\vee\left\{Trace\left (sk^*, tsk, PP\right) \neq id_*\right\}]<\epsilon (\lambda) $. 
\end{center}
\end{definition}
\subsubsection{PPTFE-IP}
A PPTFE-IP scheme comprises the same {\bf Setup,  Encryption,  Decryption, Trace} algorithms as the TFE-IP scheme mentioned in Section~\ref{fd}.  However, the {\bf KeyGen} algorithm in the TFE-IP scheme is  replaced by the {\bf PPKeyGen} algorithm. The {\bf PPKeyGen} algorithm is described below. \\

{\bf PPKeyGen}$ (User (PP, id, \vec{f}, decom_u)\!\leftrightarrow\!KGC (PP, msk, \vec{f}, com_u)\!\rightarrow\!(\bot,$\\$sk_{\vec{f},id})$.  This algorithm includes the interaction process between the user and the KGC.  Let $Commitment (PP, id) \rightarrow (com_u,  decom_u)$ be a commitment scheme which inputs the public parameters and a secret identity $id$, producing the commitment $comm_{u}$ and decommitment $decom_{u}$ as output.  Given the public parameters $PP$,   the master secret key $msk$,  a vector $\vec{f}$ and the commitment $com_u$,  KGC outputs $\perp$. The user takes the public parameters $PP$,  an identity $id $,  a vector $\vec{f}$ and $decom_u$ as input.  If $decommitment (PP, id, decom_u, com_u) =1$,  a secret key $sk_{\vec{f}, id}$ is output; Otherwise,  the output fails. 
The formalization of the security model for PPKeyGen algorithm employs two games~\cite{gh:ibe07,ckrs:ibe09}: $selective-failure-blind$ and $leakage-free$. \\

{\bf Selective-failure-blind. }The user U is honest and the KGC is malicious and in this game. KGC tries to distinguish the user's identity $id$ associated with the key. 
\begin{itemize}
\item KGC publishes public parameters $PP$ and submits two identities $id_0, id_1$. 
\item KGC  randomly selects $\delta\in\{0, 1\}$ and gets two commitments $com_\delta, com_{1-\delta}$ which belongs to $id_0$ and $id_1$ respectively.  KGC can use two oracles $U (PP, id_\delta, decom_\delta)$ and $U (PP, id_{1-\delta}, decom_{1-\delta})$ to generate $sk_\delta$ for $id_\delta$ and $sk_{1-\delta}$ for $id_{1-\delta}$. 
\begin{align*}
&1)  sk_\delta=\perp, sk_{1-\delta}=\perp,  U\ \text{returns}\  (\zeta, \zeta) \ \text{to}\ KGC;\\
&2)  sk_\delta=\perp, sk_{1-\delta}\neq\perp,  U\ \text{returns}\  (\perp, \zeta) \ \text{to}\ KGC;\\
&3)  sk_\delta\neq\perp, sk_{1-\delta}=\perp,  U\ \text{returns}\  (\zeta, \perp) \ \text{to}\ KGC;\\
&4)  sk_\delta\neq\perp, sk_{1-\delta}\neq\perp,  U\ \text{returns}\  (sk_0, sk_1) \ \text{to}\ KGC.
\end{align*}
\item KGC guesses $\delta^\prime$ about $\delta$.  In the case that $\delta^\prime=\delta$,  KGC wins the game. 
\end{itemize}
\begin{definition}[Selective-Failure-Blindness]  The PPKeyGen algorithm is selective - failure - blind if KGC is able to win the above game with negligible advantage of $\epsilon (\lambda) $, 
\begin{center}
$Adv_{KGC}=\left|\Pr[\delta^\prime=\delta]-\frac{1}{2}\right|<\epsilon (\lambda) $. 
\end{center}
\end{definition}
{\bf Leakage-free. }In this game, suppose that the user U is malicious and the KGC is honest, user U has interaction with KGC in an attempt to get informed about the key.  The above game comprises a real-world and an ideal-world scenario,  in which a distinguisher $\mathcal{D}$ tries to distinguish the outputs of each scenario. 
\begin{itemize}
\item {\bf Real-world:} User U chooses identity $\theta$ and interacts with KGC by {\bf PPKeyGen},  and $\mathcal{D}$ can see the interaction between KGC and U. 
\item {\bf Ideal-world:} The simulator $\mathcal{S}$ chooses an identity $\theta$, then asks a trusted party $TP$ to generate a key through {\bf KeyGen}. $\mathcal{D}$ witnesses the communication process between the simulator $\mathcal{S}$ and the $TP$. 
\end{itemize}
\begin{definition}[Leakage-Freeness] If $\mathcal{D}$ can distinguish real-world outputs from ideal-world outputs with only a negligible advantage of $\epsilon (\lambda) $, 
\begin{center}
$Adv_{U}=\left|\Pr[\mathcal{D} (Real_U^{PPKeyGen}) =1]-\Pr[\mathcal{D} (Ideal_\mathcal{S}^{TP}) =1]\right|<\epsilon (\lambda) $,
\end{center}
we call that the algorithm is leakage-free. \\
\end{definition}

\begin{definition}[Security of PPTFE-IP]
A PPTFE-IP scheme $\Omega=({\bf Setup,}$\\${\bf Encrypt, PPKeyGen, Decrypt, Trace}) $ is secure if the following two conditions are satisfied:
\begin{itemize}
\item The TFE-IP scheme $\Delta\!=\!\left({\bf Setup, Encrypt, KeyGen, Decrypt,Trace}\right)$ is s-IND-CPA secure;
\item The {\bf PPKeyGen} algorithm satisfies two properties: selective-failure-blindness and leakage-freeness. 
\end{itemize}
\end{definition}
\section{Our Constructions}\label{sec3}

Firstly, a traceable FE-IP scheme is concretely constructed firstly, then we present the construction of the {\bf PPKeyGen} algorithm.

\subsection{Our TFE-IP Scheme}\label{subsec2}
An overview of our TFE-IP scheme are is as follows.\\ 
\begin{figure}[!t]
\begin{framed}
\scriptsize
{\bf Setup}.~Suppose that $\mathcal{BG}(1^\lambda ) \rightarrow (e, p, \mathbb{G}, \mathbb{G}_T)$ and  $g0, g1, g2\in\mathbb{G}$ are generators. KGC selects $\vec{s}= (s_1, s_2,\cdots, s_l) \stackrel{R}{\leftarrow}\mathbb{Z}_p^l$ and computes $\vec{h}=\{h_i=g_1^{s_i}\}\ for\ i \in [l]$. Tracer selects $b \stackrel{R}{\leftarrow}\mathbb{Z}_p$ randomly, calculating $B=g_2^b$. KGC selects a random $a$ and publishes $Y=g_0^a$. KGC sets $msk=(a,s_{i})$ as master secret key, publishing $ (\vec{h}, Y) $. The tracer's secret-public key pair is $(b, B)$. Public parameters of the system can be denoted as $PP= (e, p, \mathbb{G}, \mathbb{G}_T, g_0, g_1, g_2, B, Y, h_1, \cdots ,h_l)$.  

{\bf Encrypt}. To encrypt a vector $\vec{x}= (x_1, x_2,\cdots, x_l) \in\mathbb{Z}_p^l$, the data owner first chooses an $r \stackrel{R}{\leftarrow}\mathbb{Z}_p$ randomly, then computes \begin{center}$ct_i=h_i^r \cdot g_1^{x_i}\ for\ i \in [l], ct_{l+1}=g_1^r, ct_{l+2}=g_2^r, ct_{l+3}=g_0^r$, \end{center}
The ciphertext is $Ct=\left ({ (ct_i) }_{i \in [l]}, ct_{l+1}, ct_{l+2}, ct_{l+3}\right) $. 

{\bf KeyGen}.  Given a vector $\vec{y} = (y_1, y_2,\cdots, y_l)$, to generate a secret key for a user with an identity $\theta$, KGC chooses random numbers $w, d\in\mathbb{Z}_{p}$ and computes
\begin{center}$K_1=g_0^{\langle \vec{y}, \vec{s}\rangle}\cdot B^{\frac{w}{d+a}}, K_2= \left (g_0 \cdot (g_2\cdot B) ^w \cdot g_2^\theta\right) ^{\frac{1}{d+a}}, K_{3}=g_1^{\frac{1}{d+a}}, K_4=w, K_5={d}$. 
\end{center}
The user receives the key $sk_{\vec{y}, \theta}$ and verifies
\begin{itemize}
\scriptsize
\item $e\left (K_1, g_1\right) = e\left (g_0, \left (\prod \limits_{i=1}^l (h_i) ^{y_i}\right) \right) \cdot e (B^w, K_3) $;
\item $e\left (K_3, g_0^{K_5}\cdot Y\right) =e\left (g_0, g_1\right) $;
\item $e\left (K_2, g_0^{K_5}\cdot Y\right) {=}e\left (g_0, g_0\right) \cdot e\left (g_0, g_2\cdot B\right) ^{K_4}\cdot e\left (g_0, g_2\right) ^\theta$. 
\end{itemize}\leavevmode
If all the above equations hold, the secret key $sk_{\vec{y}, \theta}$ is valid; otherwise, it is invalid.

{\bf Decrypt}. A user uses his/her secret key $(K_1,K_2,K_3,K_4,K_5)$ to decrypt the ciphertext as follows:
\begin{center}
$e (g_0, g_1) ^{\langle\vec{x}, \vec{y}\rangle}=\frac{e\left (g_0, \prod \limits_{i=1}^l  (ct_i) ^{y_i}\right) \cdot e (ct_{l+1}, K_2) }{ e (K_1, ct_{l+1}) \cdot e (K_3, ct_{l+3}) \cdot e (K_3^{K_4} \cdot K_3^\theta, ct_{l+2}) }$
\end{center}\leavevmode\\
Further, the user calculates the discrete logarithm of $e (g_0, g_1) ^{\langle\vec{x}, \vec{y}\rangle}$ with respect to $e (g_0, g_1)$ to obtain ${\langle\vec{x}, \vec{y}\rangle}$. This discrete logarithmic operation requires that ${\langle\vec{x}, \vec{y}\rangle}$ should not be too large.

{\bf Trace}. Given a key $sk_{\vec{y}, \theta}$, the tracer can compute
\begin{center} $e\left (K_3, g_2\right) ^\theta=\frac{e (K_2, g_1) }{e (g_0, K_3) \cdot e (g_2,  K_3^{K_4} \cdot K_3^{{K_4} \cdot b}) }$ \end{center}
to recover the user's identity associated with the key, where $b$ denotes for tracer's secret key.
\end{framed}
\caption{Our TFE-IP Scheme}\label{fig2}
\end{figure}
The proposed TFE-IP scheme proceeds as below:
\begin{itemize}
\item Firstly, system runs ${\bf Setup}$, KGC generates a master secret key and public parameters. The tracer calculates a secret-public key pair.
\item Secondly, for the purpose of computing the inner product of $\vec{x}$ and $\vec{y}$ , the user needs to get the key $sk_{\vec{y}, \theta}=\{K_1,K_2,K_3,K_4,K_5\}$ with the identity $\theta$ and a vector $\vec{y}$ embedded from KGC and the ciphertext of a vector $\vec{x}$ from data owner. In particular, $K_1$ is bound with the vector $\vec{y}$ and $K_2$ is bound with the identity $\theta$. KGC binds the identity and vector together in users' secret keys using a random number. $K_3, K_4$ and $K_5$ are used in tracing and decryption. To prevent collusion attacks, all elements included in a key are bound together by a random number. Users are able to verify the correctness of his/her secret key.
\item Thirdly, a vector is encrypted by the data owner using the public parameters.
\item Fourthly,  given a ciphertext, user's secret key can be utilized by the user to compute the inner-product of two vectors respectively associated with the ciphertext and his/her secret key. Additionally,  $e (g_0, g_1) ^{\langle\vec{x}, \vec{y}\rangle}$  needs to be small to solve ${\langle\vec{x}, \vec{y}\rangle}$.
\item Finally, if tracing is required, only tracer can use the trace secret key to recover $e\left (K_3, g_2\right) ^\theta$ from the secret key $sk_{\vec{y}, \theta}$. Tracer computes the discrete logarithm of each identity based on $e\left (K_3, g_2\right)$ to discover the identity embedded in the secret key.
\end{itemize}

{\it Correctness of our TFE-IP Scheme}.  The correctness of our TFE-IP scheme is shown by the following equations.
\begin{align*}
&e\left (g_0, \prod \limits_{i=1}^l  (ct_i) ^{y_i}\right)\!\cdot\! e (ct_{l+1}, K_2)\!=\!\left (\prod\limits_{i=1}^l e (g_0, h_i^r \!\cdot\! g_1^{x_i}) ^{y_i}\right)\!\cdot\! e (g_1^r, \left (g_0\!\cdot\! (g_2\!\cdot\! B) ^w \!\cdot\! g_2^\theta\right) ^{\frac{1}{d+a}}) \\
&=e (g_0, g_1) ^{\langle\vec{x}, \vec{y}\rangle}\cdot e (g_0, g_1^r) ^{\langle \vec{y}, \vec{s}\rangle}\cdot e (g_1^{\frac{1}{d+a}}, g_0^r) \cdot e (g_1^{\frac{w}{d+a}}, g_2^r)  \cdot e (g_1^{\frac{w}{d+a}}, B^r) \cdot e (g_1^{\frac{1}{d+a}}, g_2^r) ^\theta\\
&=e (g_0, g_1) ^{\langle\vec{x}, \vec{y}\rangle}\cdot e (g_0, g_1^r) ^{\langle \vec{y}, \vec{s}\rangle}\cdot e (g_1^{\frac{w}{d+a}}, B^r) \cdot e (g_1^{\frac{1}{d+a}}, g_0^r) \cdot e (g_1^{\frac{w}{d+a}}, g_2^r) \cdot e (g_1^{\frac{1}{d+a}}, g_2^r) ^\theta\\
&=e (g_0, g_1) ^{\langle\vec{x}, \vec{y}\rangle}\cdot e (g_0^{\langle \vec{y}, \vec{s}\rangle}\cdot B^{\frac{w}{d+a}}, g_1^r) \cdot e (g_1^{\frac{1}{d+a}}, g_0^r) \cdot e (g_1^{\frac{w}{d+a}}, g_2^r) \cdot e (g_1^{\frac{1}{d+a}}, g_2^r) ^\theta\\
&=e (g_0, g_1) ^{\langle\vec{x}, \vec{y}\rangle}\cdot e (K_1, ct_{l+1}) \cdot e (K_3, ct_{l+3}) \cdot e (K_3, ct_{l+2}) ^w\cdot e (K_3, ct_{l+2}) ^\theta\\
&=e (g_0, g_1) ^{\langle\vec{x}, \vec{y}\rangle}\cdot e (K_1, ct_{l+1}) \cdot e (K_3, ct_{l+3}) \cdot e (K_3^{K_4} \cdot K_3^\theta, ct_{l+2}),
\end{align*}
and\ $K_1\!=\!g_0^{\langle \vec{y}, \vec{s}\rangle}\cdot B^{\frac{w}{d+a}}, K_2\!= \!\left (g_0\cdot (g_2\cdot B) ^w \cdot g_2^\theta\right) ^{\frac{1}{d+a}}, K_{3}\!=\!g_1^{\frac{1}{d+a}}, K_4\!=\!w, K_5\!=\!{d}$.\\
Therefore, $\frac{e\left (g_0, \prod \limits_{i=1}^l  (ct_i) ^{y_i}\right) \cdot e (ct_{l+1}, K_2) }{ e (K_1, ct_{l+1}) \cdot e (K_3, ct_{l+3}) \cdot e (K_3^{K_4} \cdot K_3^\theta, ct_{l+2}) }=e (g_0, g_1) ^{\langle\vec{x}, \vec{y}\rangle}$.\\
For\ tracing, $\frac{e (K_2, g_1) }{e (g_0, K_3) \cdot e (g_2,  K_3^{K_4} \cdot K_3^{{K_4} \cdot b}) }=\frac{e ( (g_0\cdot (g_2\cdot B) ^w\cdot g_2^\theta) ^{\frac{1}{d+a}}, g_1) }{e (g_0, g_1^{\frac{1}{d+a}}) \cdot e (g_2, g_1^{\frac{w}{d+a}} \cdot  (g_1^{\frac{w}{d+a}}) ^b) }=e\left (K_3, g_2\right) ^\theta$.
\subsection{Our PPKeyGen Algorithm}
In order to prevent user collusion attacks, user identity is associated with his/her secret key. Therefore,the identity of each key holder and his/her secret key are known to the KGC. Considering privacy issues, we introduce a {\bf PPKeyGen} algorithm in which each user and the KGC collaborate to generate a secret key using secure two-party computing.   The user and the KGC collaborate to generate the key in {\bf PPKeyGen} algorithm, while other users and the KGC cannot learn the embedded identity from the key. Figure 3 shows the construction of our {\bf PPKeyGen} algorithm. The instantiation of zero-knowledge proof $\Sigma_K$ and $\Sigma_U$ in our {\bf PPKeyGen} algorithm are presented in~\ref{app}. 

\begin{table}
\scriptsize
\newcolumntype{Y}{>{\raggedright\arraybackslash}X}
\newcolumntype{Z}{>{\centering\arraybackslash}X} 
\begin{tabularx}{\linewidth}{|Yp{2cm}Y|}
\hline
 & {\bf PPKeyGen} &  \\
\bf User & & {\bf KGC} \\
$\left (GID\ \theta, w_1\stackrel{R}{\leftarrow}\mathbb{Z}_p, \tau\stackrel{R}{\leftarrow}\mathbb{Z}_p\right) $ &              &$ \left (msk\  \vec{s}, w_2, d\stackrel{R}{\leftarrow}\mathbb{Z}_p\right) $ \\
1. Select\ $w_1\stackrel{R}{\leftarrow}\mathbb{Z}_p, \tau\stackrel{R}{\leftarrow}\mathbb{Z}_p$, and compute&              &  \\
${A}_{1}=h^{\tau}\cdot {B}^{w_1}, A_{2}={ (g_2\cdot B) }^{w_1}\cdot g_2^\theta, $&             &2. Select\ $w_2\stackrel{R}{\leftarrow}\mathbb{Z}_p$\ and\ compute \\
Generate\ $\Sigma_U=PoK\left\{\left (w_1, \theta, \tau\right) :\right.$&    $\xrightarrow[\Sigma_U]{A_1, A_2}$     & $B_1=\prod \limits_{i=1}^l  (g_0^{y_i}) ^{s_i}\cdot  (A_1\cdot B^{w_2}) ^{\frac{1}{d+a}}$, \\
$\left.{A}_{1}=h^{\tau}\cdot {B}^{w_1}\wedge A_{2}={ (g_2\cdot B) }^{w_1}\cdot g_2^\theta\right\}$&		    &$B_2= (g_0 \cdot A_2\cdot  (g_2\cdot B) ^{w_2}) ^{\frac{1}{d+a}}$, \\
&              &$B_3=g_1^{\frac{1}{d+a}}, B_4=h^{\frac{1}{d+a}}, B_5=d$.  \\
&            &Generate\ $\Sigma_K=PoK\left\{\left (a, w_{2},  { (s_i) }_{i \in [l]}\right) :\right.$\\
3. Compute $w=w_1+w_2$ and set&            &$B_2= (g_0 \cdot A_2\cdot  (g_2\cdot B) ^{w_2}) ^{\frac{1}{d+a}}\wedge  $\\
 $K_1=\frac{B_1}{B_4^\tau}, K_2=B_2, K_3=B_3,$ &           $\xleftarrow[B_4, B_5, \Sigma_K]{w_2, B_1, B_2, B_3}$   &  $B_1=g_0^{\langle\vec{y}, \vec{s}\rangle}\cdot A_1^{\frac{1}{d+a}} \cdot B^{\frac{w_2}{d+a}}\wedge$\\
$K_4=w, K_5=B_5$&		    &$\left.B_3=g_1^{\frac{1}{d+a}}\wedge B_4=h^{\frac{1}{d+a}}\right\}$\\
\hline
\end{tabularx}
\caption*{Figure 3:  Our {\bf PPKeyGen} algorithm}
\end{table}

{\it Correctness of Our Privacy-Preserving Key Generation Algorithm}.  Let $w=w_1+w_2$, the equations presented below demonstrate the correctness of the secret keys generated in Figure3. 
\begin{align*}
K_1&=\frac{B_1}{B_4^\tau}=\frac{g_0^{\langle\vec{y}, \vec{s}\rangle}\cdot A_1^{\frac{1}{d+a}} \cdot B^{\frac{w_2}{d+a}}}{\left (h^{\frac{1}{d+a}}\right) ^\tau}=\frac{B_1}{B_4^\tau}=\frac{g_0^{\langle\vec{y}, \vec{s}\rangle}\cdot \left (h^{\tau}\cdot {B}^{w_1}\right) ^{\frac{1}{d+a}} \cdot B^{\frac{w_2}{d+a}}}{\left (h^{\frac{1}{d+a}}\right) ^\tau}\\
&=g_0^{\langle \vec{y}, \vec{s}\rangle}\cdot B^{\frac{w_1+w_2}{d+a}}=g_0^{\langle \vec{y}, \vec{s}\rangle}\cdot B^{\frac{w}{d+a}},\\
K_2&=B_2= (g_0 \cdot A_2\cdot  (g_2\cdot B) ^{w_2}) ^{\frac{1}{d+a}}= (g_0 \cdot \left ({ (g_2\cdot B) }^{w_1}\cdot g_2^\theta\right) \cdot  (g_2\cdot B) ^{w_2}) ^{\frac{1}{d+a}}\\
&=\left (g_0 \cdot (g_2\cdot B) ^{ (w_1+w_2) } \cdot g_2^\theta\right) ^{\frac{1}{d+a}}=\left (g_0 \cdot (g_2\cdot B) ^w \cdot g_2^\theta\right) ^{\frac{1}{d+a}},\\
K_3&=g_1^{\frac{1}{d+a}},K_4=w_1+w_2=w,K_5=d.
\end{align*}

\section{Security Analysis}\label{sec4}
\theorem{Our TFE-IP scheme is $(\epsilon (\lambda), t)$ secure against chosen-plaintext attack (CPA) in the selective model if the DDH assumption
holds on the group $\mathbb{G}$ with $ (\epsilon (\lambda)^\prime,t^\prime)$, where $\epsilon (\lambda)^\prime =\frac{\epsilon (\lambda) }{2}$. }\label{ind-cpa}
\begin{proof}[Proof] Suppose the existence of an adversary $\mathcal{A}$ who can $ (t, \epsilon)$-break the TFE-IP scheme in the IND-CPA security model, there exists a simulator $\mathcal{B}$ who can run $\mathcal{A}$ to break the $DDH$ assumption as below.  Challenger $\mathcal{C}$ randomly selects a $\mu \in \{0, 1\}$.  If $\mu=0$,  $\mathcal{C}$ sends $ (g^\alpha, g^\beta, Z=g^{\alpha\beta}) $ to $\mathcal{B}$; if $\mu=1$, $\mathcal{C}$ sends  $ (g^\alpha, g^\beta, Z=g^{\tau}) $ to $\mathcal{B}$,  in which $\tau\in \mathbb{Z}_p$ is a random number.  $\mathcal{B}$ outputs its guess $\mu^\prime$ about $\mu$.  
\begin{itemize}
\item {\bf Initialization.} $\mathcal{A}$ submits two vectors $\vec{x_0}= (x_{0, 0}, x_{0, 1},\cdots x_{0, l}), \vec{x_1}= (x_{0, 0}, x_{0, 1},\cdots,x_{0, l})$, in which $l$ represents the length of vectors. 
\item {\bf Setup.} $\mathcal{B}$ selects random $a, c_0,c_2\leftarrow \mathbb{Z}_p$, setting $g_1=g, g_0=g^{c_0}, g_2=g^{c_2}, Y=g_0^a$.  Let $spc (\vec{x_0}-\vec{x_1}) $ be a linear space with basis $ (\vec{\eta_1}, \vec{\eta_2},\cdots, \vec{\eta_l}) $ and the basis of $spc (\vec{x_0}-\vec{x_1}) ^\perp$ is $ (\vec{\zeta_1}, \vec{\zeta_2},\cdots, \vec{\zeta_l}) $. $\mathcal{B}$ randomly selects a vector $\vec{\pi}= ({\pi_1}, {\pi_2},\cdots, {\pi_l}) \in spc (\vec{x_0}-\vec{x_1}) ^\perp$,  and computes $ (h_i= (g_1^\alpha) ^{\pi_i}) _{i\in[l]}$.   $\mathcal{B}$ returns public parameters $PP=\left\{g_0,g_1,g_2,Y, (h_i) _{i\in[l]}\right\}$ to $\mathcal{A}$. $\mathcal{B}$ impliedly defined $\vec{s}= (s_i) _{i\in[l]}=  (\alpha\cdot\pi_i) _{i\in[l]}$.                
\item {\bf Phase-I.} $\mathcal{A}$ submits $\vec{y}= (y_i) _{i\in[l]}\in \mathbb{Z}_p, \theta \in \mathbb{Z}_p$ with the limitation that $\vec{y}\in spc (\vec{x_0}-\vec{x_1}) ^\perp$.  $\mathcal{B}$ selects random $w, d \stackrel{R}{\leftarrow}\mathbb{Z}_p$ and computes
\begin{center}$K_1=g_0^{\langle \vec{y}, \vec{s}\rangle}\cdot B^{\frac{w}{d+a}}=B^{\frac{w}{d+a}}, K_2= \left (g_0 \cdot (g_2\cdot B) ^w \cdot g_2^\theta\right) ^{\frac{1}{d+a}}, K_{3}=g_1^{\frac{1}{d+a}}, K_4=w, K_5={d}$. 
\end{center}
$\mathcal{B}$ returns to $\mathcal{A}$ $sk_{\vec{y}, U}=\left ({ (K_i) }_{i \in [5]}\right) $. \\
\item {\bf Challenge.} $\mathcal{B}$ randomly chooses a $\delta \in \{0, 1\}$.  $\mathcal{B}$ calculates
\begin{center} 
$ct_i^*=Z^{\pi_i} \cdot g^{x_{\delta, i}}\ for\ i \in [l], ct_{l+1}^*=g^\beta, ct_{l+2}^*={g^\beta}^{c_2}, ct_{l+3}^*={g^\beta}^{c_0}$, 
\end{center}
outputs ciphertext $Ct^*=\left ({ (ct_i^*) }_{i \in [l]}, ct^*_{l+1}, ct^*_{l+2}, ct^*_{l+3}\right) $. 
\item {\bf Phase-II.} Repeat the same process as in {\bf Phase-I}. 
\item {\bf Output.} $\mathcal{A}$ guesses $\delta^\prime$ about $\delta$.  If\ $\delta^{\prime}\neq\delta,  \mathcal{B}$\ produces\ $\mu^\prime=1$ as output; If\ $\delta^{\prime} =\delta, \mathcal{B}$\ produces\ $\mu^\prime=0$ as output.
\end{itemize}
The remaining thing to compete the proof is to calculate the advantage with which $\mathcal{B}$ can break the DDH assumption.
\begin{itemize}
\item If $\mu=0, Z=g^{\alpha\beta}, ct_i^*= Z^{\pi_i} \cdot g^{x_{\delta, i}}={g^{\alpha\beta}}^{\pi_i} \cdot g^{x_{\delta, i}}, ct_{l+1}^*=g^\beta, ct_{l+2}^*={g^\beta}^{c_2}, ct_{l+3}^*={g^\beta}^{c_0}$ is a correct ciphertext of $x_\delta$.  Therefore,  $\Pr[\delta^\prime=\delta|\mu=0]=\frac{1}{2}+\epsilon (\lambda) $. When $\delta^{\prime} =\delta, \mathcal{B}\ $outputs\ $\mu^\prime=0$, so $\Pr[\mu^\prime=\mu|\mu=0]=\frac{1}{2}+\epsilon (\lambda) $. 
\item If $\mu=1, Z=g^{\tau}, ct_i^*= Z^{\pi_i} \cdot g^{x_{\delta, i}}={g^{\tau}}^{\pi_i} \cdot g^{x_{\delta, i}}, ct_{l+1}^*=g^\beta, ct_{l+2}^*={g^\beta}^{c_2}, ct_{l+3}^*={g^\beta}^{c_0}$.  This information theoretically hide both $x_0$ and $x_1$.  Therefore,  $\Pr[\delta^\prime\neq\delta|\mu=1]=\frac{1}{2}$. When $\delta^{\prime}\neq\delta,  \mathcal{B}$\ produces\ $\mu^\prime=1$ as output,  so $\Pr[\mu^\prime=\mu|\mu=1]=\frac{1}{2}$. 
\end{itemize}
In conclusion,  the advantage of breaking the $DDH$ assumption by $\mathcal{B}$ can be computed as follows:
\begin{center}
$\left|\frac{1}{2}\Pr[\mu^\prime=\mu|\mu=0]- \frac{1}{2}\Pr[\mu^\prime\neq\mu|\mu=1]\right|=\frac{1}{2}\cdot\left (\frac{1}{2}+\epsilon (\lambda) \right) -\frac{1}{2}\cdot\frac{1}{2}=\frac{\epsilon (\lambda) }{2}$. 
\end{center}
\end{proof}
\subsection{Traceability}
{\theorem The proposed TFE-IP scheme is $ (\delta, \epsilon) -traceable$ if the q-SDH assumption holds with an advantage no greater than $\epsilon_1$,  and the DL assumption holds with an advantage no greater than $\epsilon_2$, where $\epsilon= \max\left\{\frac{\epsilon_1}{4}\left(1-\frac{q}{p}\right)+\frac{\epsilon_2}{4}\frac{p-1}{p^3},\right.$\\$\left. \frac{\epsilon_1}{4}\left(\left(1-\frac{q}{p}\right)+\frac{1}{q}\right)\right\}$ and $\delta< q$ represents the number of queries made by $\mathcal{A}$. }\label{trac}
\begin{proof}[Proof] Suppose the existence of an adversary $\mathcal{A}$, a simulator $\mathcal{B}$ who is able to run $\mathcal{A}$ to solve the SDH problem as follows. The challenger $\mathcal{C}$ sends $\left(g, g^{z},g^{z^{2}},\cdots, g^{z^{q}}, h, h^{z}\right)$ to $\mathcal{B}$. $\mathcal{B}$ will produce $(c,g^{z+c})$ as output where $c\in \mathcal{Z}_{p}$
\begin{itemize}
\item {\bf Setup.} $\mathcal{B}$ sets $ (\Phi_i=g^{z_i}) _{i\in[q]}$, randomly selects $a, \mu_1, \mu_2,\cdots \mu_{q-1}\stackrel{R}{\leftarrow}\mathbb{Z}_p$, $\vec{s}= (s_1, s_2,\cdots, s_l) \stackrel{R}{\leftarrow}\mathbb{Z}_p^l$,  and lets $f (z) ={\prod_{i=1}^{q-1}  (z+\mu_i) }={\Sigma_{i=0}^{q-1} {\delta_i} z^i}$, then computes $\tilde{g}=\prod_{i=0}^{q-1} (g^{z_i}) ^{\delta_i}=g^{f (z) }, \hat{g}=\prod_{i=0}^{q-1} (g^{z^{i+1}}) ^{\delta_i}=\tilde{g}^z, \vec{h}=\{h_i=g_1^{s_i}\}\ for\ i \in [l], Y=g_0^a$. 
$\mathcal{B}$ chooses $\gamma_1, \gamma_2, \gamma_3, \pi, \rho, \kappa, b$ $\stackrel{R}{\leftarrow}\mathbb{Z}_p$, sets $B=\tilde{g}^b, $ then calculates $g_0=\tilde{g}^{\gamma_1}, g_2=\left (\left (\hat{g}\tilde{g}^\pi\right) ^\kappa\tilde{g}^{-1}\right) ^{\frac{1}{\rho}}=\tilde{g}^{\frac{ (z+\pi) \kappa-1}{\rho}}, \tilde{g}=g^{\gamma_3}$.  $B$ is the trace public key.  $\mathcal{A}$ is given public parameters $ (e, p, \tilde{g}, \hat{g}, g_0, g_1, g_2, g_3, B, Y, h_1, \cdots ,h_l) $ from $\mathcal{B}$. \\
\item {\bf Key Query.}  $\mathcal{A}$ sends $ (\theta_i, \vec{y}_i) _{i\leq q}$ to $\mathcal{B}$ for the $i-th$ query. $\mathcal{B}$ sets $f_i (z) =\frac{f (z) }{z+\mu_i}={\Sigma_{j=0}^{q-2} {\eta_i}_j z^j}$, randomly selects $w\stackrel{R}{\leftarrow}\mathbb{Z}_p$, computes  
\begin{align*}
\nu_i&=\prod_{j=0}^{q-2} (\Phi_j) ^{{\eta_i}_j}={g}^{f_i (z) }=g^{\frac{f (z) }{z+\mu_i}}=\tilde{g}^{\frac{1}{z+\mu_i}},\\
K_1&=g_0^{\langle \vec{y}, \vec{s}\rangle}\cdot\nu_i^{b\cdot w},K_3=\nu_i^{\gamma_1}, K_4=w, K_5=\mu_i, \\
K_2&=\prod_{j=0}^{q-2}\left (g^ {z^{j+1}}\right) ^{{{\eta_i}_j}\cdot {\frac{\kappa}{\rho}} \cdot (\gamma_2\cdot w+\theta_i) }\cdot\prod_{j=0}^{q-2}\left (g^ {z^{j}}\right) ^{{\eta_i}_j\cdot\left ( {\left (\frac{ (\pi\cdot\kappa-1) }{\rho}\right) } \cdot (\gamma_2\cdot w+\theta_i) +\gamma_1\right) },
\end{align*}
and returns $sk_{ (\theta_i, \vec{y}_i) }\!=\!\left (K_1,K_2,K_3,K_4,K_5\right) $ to $\mathcal{A}$. We prove that $sk_{ (\theta_i, \vec{y}_{i}) }$ is correct. 
\begin{align*}
K_1&=g_0^{\langle \vec{y}, \vec{s}\rangle}\cdot\nu_i^{b\cdot w}=g_0^{\langle \vec{y}, \vec{s}\rangle}\cdot \tilde{g}^{\frac{b \cdot w}{z+\mu_i}}=g_0^{\langle \vec{y}, \vec{s}\rangle}\cdot B^{\frac{w}{z+\mu_i}}, \\
K_2&=\prod_{j=0}^{q-2}\left (g^ {z^{j+1}}\right) ^{{{\eta_i}_j}\cdot {\frac{\kappa}{\rho}} \cdot (\gamma_2\cdot w+\theta_i) }\cdot\prod_{j=0}^{q-2}\left (g^ {z^{j}}\right) ^{{\eta_i}_j\cdot\left ( {\left (\frac{ (\pi\cdot\kappa-1) }{\rho}\right) } \cdot (\gamma_2\cdot w+\theta_i) +\gamma_1\right) }, \\
&=\prod_{j=0}^{q-2}g^{{{\eta_i}_j\cdot z^{j+1}}\cdot {\frac{\kappa}{\rho}} \cdot (\gamma_2\cdot w+\theta_i) }\cdot\prod_{j=0}^{q-2}g^{{{\eta_i}_j\cdot z^{j}}\cdot\left ( {\left (\frac{ (\pi\cdot\kappa-1) }{\rho}\right) } \cdot (\gamma_2\cdot w+\theta_i) ++\gamma_1\right) }, \\
&=g^{{ ({\Sigma_{j=0}^{q-2}{\eta_i}_j\cdot z^{j+1}) }\cdot {\frac{\kappa}{\rho}} \cdot (\gamma_2\cdot w+\theta_i) } }\cdot g^{ (\Sigma_{j=0}^{q-2}{\eta_i}_j\cdot z^{j}) \cdot\left (\left (\frac{ (\pi\cdot\kappa-1) }{\rho}\right)  \cdot{ (\gamma_2\cdot w+\theta_i) }+\gamma_1\right) }\\
&=g^{{ (\Sigma_{j=0}^{q-2}{\eta_i}_j\cdot z^j) }\cdot\left ({\frac{ ( (z+\pi) \kappa-1) \cdot (\gamma_2\cdot w+\theta_i) }{\rho}}+\gamma_1\right) }\\
&={g}^{\gamma_1\cdot f_i (z) }\cdot g^{f_i (z) \cdot {\frac{ (z+\pi) \kappa-1}{\rho}} \cdot{ (\gamma_2\cdot w+\theta_i) }}=\tilde{g}^{\frac{\gamma_1}{z+\mu_i}}\cdot\tilde{g}^{\frac{ (z+\pi) \kappa-1}{\rho}\cdot{\frac{\gamma_2\cdot w+\theta_i}{z+\mu_i}}}\\
&=g_0^{\frac{1}{z+\mu_i}}\cdot g_2^{\frac{\gamma_2\cdot w+\theta_i}{z+\mu_i}}= (g_0\cdot (g_2 \cdot B) ^w\cdot g_2^{\theta_i}) ^{\frac{1}{z+\mu_i}}, \\
K_3&=g_1^{\frac{1}{z+\mu_i}}=\tilde{g}^{\frac{\gamma_1}{z+\mu_i}}=\nu_i^{\gamma_1}. 
\end{align*}
\item {\bf Trace Query.} $\mathcal{A}$ adaptively submit a key  $sk_{i}= ({K}_1, {K}_2, {K}_3, {K}_4, {K}_5) $ to $\mathcal{B}$.  $\mathcal{B}$ computes
\begin{center} $\frac{e (K_2, g_1) }{e (g_0, K_3) \cdot e (g_2,  K_3^{K_4} \cdot K_3^{{K_4} \cdot b}) }=e\left (K_3, g_2\right) ^\theta$ \end{center}\leavevmode\\
$\mathcal{B}$ sends to $\mathcal{A}$ the discrete logarithm of identity $\theta$. 
\item {\bf Key Forgery.} $\mathcal{A}$ $sk_{ (\theta^*, \vec{y}^*) }=\left (K_1^*, K_2^*, K_3^*, K_4^*, K_5^*\right) $ and sends the key to $\mathcal{B}$. Let's consider the two types of forgery as follows.
\item {\bf Type I} The identity associated with the key hasn't been previously queried,  namely,  $\theta^*\notin\{\theta_1, \theta_2,\cdots \theta_q\}$. Furthermore, this forgery is divided into two cases: \\
$\bm{Case-I:\ \theta^*\notin\{\theta_1, \theta_2,\cdots \theta_q\}, \mu^*\notin\{\mu_1, \mu_2,\cdots \mu_{q-1}, \pi\}}$. \\
Set $f (z) ^*_1=\frac{f (z) }{z+\mu_i}={\Sigma_{j=0}^{q-2} {\psi}_j z^j}, f (z) ^*_2=\frac{f (z) \cdot (z+\pi) }{z+\mu_i}={\Sigma_{j=0}^{q-1} {\psi^{\prime}}_j z^j}, f (z) = (z+\mu^*) \cdot \sigma (z) +\chi, \sigma (z) =\Sigma_{j=0}^{q-2} \sigma_j\cdot z^j$ and thus ${K_2^*}= (g_0\cdot (g_2 \cdot B) ^{K_4^*}\cdot g_2^{\theta^*}) ^{\frac{1}{z+K_5^*}}= (g_0\cdot (g_2 \cdot B) ^{w^*}\cdot g_2^{\theta^*}) ^{\frac{1}{z+\mu^*}}= (g_0^{\frac{1}{z+\mu^*}}) \cdot ( (g_2 \cdot B) ^{w^*}\cdot g_2^{\theta^*}) ^{\frac{1}{z+\mu^*}}.$ Furthermore, we have
\begin{align*}
&g_0^{\frac{1}{z+\mu^*}}\!=\!{K_2^*}\!\cdot\! ( (g_2 \!\cdot\! B) ^{w^*}\!\cdot\! g_2^{\theta^*}) ^{\frac{-1}{z\!+\!\mu^*}}\!=\!{K_2^*}\!\cdot\!( (\tilde{g}^{\frac{ ( (z\!+\!\pi) \kappa-1)\!\cdot\!{\gamma_2\!\cdot\! w^*}}{\rho}}) \!\cdot\! \tilde{g}^{\frac{ ( (z+\pi) \kappa-1) \!\cdot\!{\theta^*}}{\rho}}) ^{\frac{-1}{z+\mu^*}}\\
&\!=\!{K_2^*}\!\cdot\!\tilde{g}^{\frac{- ( (z+\pi) \kappa)\!\cdot\!{ (\gamma_2\cdot w^*+\theta^*) }}{\rho\!\cdot\!(z+\mu^*) }}\!\cdot\!\tilde{g}^{\frac{\gamma_2\!\cdot\!w^*\!+\!\theta^*}{\rho\!\cdot\!(z\!+\!\mu^*) }}\!=\!{K_2^*}\!\cdot\!{g}^{\frac{-f^ (z)\!\cdot\!( (z+\pi) \kappa)\!\cdot\!{ (\gamma_2\cdot w^*+\theta^*) }}{\rho\cdot (z+\mu^*) }}\cdot{g}^{\frac{f^ (z) \cdot (\gamma_2\cdot w^*+\theta^*) }{\rho\cdot (z+\mu^*) }}\\
&={K_2^*}\cdot{g}^{\frac{-f^*_2 (z) \cdot\kappa) \cdot{ (\gamma_2\cdot w^*+\theta^*) }}{\rho}}\cdot{g}^{\frac{f^*_1 (z) \cdot (\gamma_2\cdot w^*+\theta^*) }{\rho}}\\
&={K_2^*}\cdot\prod_0^{q-1}{\left (g^{z^i}\right) }^{\frac{-\psi^{\prime}_i\cdot\kappa\cdot{ (\gamma_2\cdot w^*+\theta^*) }}{\rho}}\cdot\prod_0^{q-2}{\left (g^{z^i}\right) }^{\frac{\psi_i\cdot (\gamma_2\cdot w^*+\theta^*) }{\rho}}. 
\end{align*}
By setting $\Theta={K_2^*}\cdot\prod_0^{q-1}{\left (g^{z^i}\right) }^{\frac{-\psi^{\prime}_i\cdot\kappa\cdot{ (\gamma_2\cdot w^*+\theta^*) }}{\rho}}\cdot\prod_0^{q-2}{\left (g^{z^i}\right) }^{\frac{\psi_i\cdot (\gamma_2\cdot w^*+\theta^*) }{\rho}}$, we have
\begin{center}
$\Theta=g_0^{\frac{1}{z+\mu^*}}=g^{\frac{f (z) \cdot\gamma_1}{z+\mu^*}}={g^{\gamma_1}}^{\frac{ (z+\mu^*) \cdot \sigma (z) +\chi}{z+\mu^*}}={g^{\gamma_1}}^{\sigma (z) }\cdot {g^{\gamma_1}}^{\frac{\chi}{z+\mu^*}}$. 
\end{center}
Therefore, we have
\begin{align*}
&g^{\frac{1}{z+\mu^*}}=\left (\Theta\cdot g^{-\sigma (z) }\right) ^{\frac{1}{\chi\cdot{\gamma_1}}}\\
&\!=\!\left ({K_2^*}\!\cdot\!\prod_{i=0}^{q-1}{\left (g^{z^i}\right) }^{\frac{-\psi^{\prime}_i\!\cdot\!\kappa\!\cdot\!{ (\gamma_2\!\cdot\!w^*+\theta^*) }}{\rho}}\!\cdot\!\prod_{j=0}^{q-2}{\left (g^{z^j}\right) }^{\frac{\psi_j\cdot (\gamma_2\!\cdot\!w^*+\theta^*) }{\rho}}\!\cdot\!\prod^{q-2}_{k=0}{\left (g^{z^k}\right) ^{ (-\sigma_k) }}\right) ^{\frac{1}{\chi\!\cdot\!{\gamma_1}}}
\end{align*}
$\mathcal{B}$ can output $ (\mu^*, g^{\frac{1}{z+\mu^*}}) $ from the above generation. Therefore,  $\mathcal{B}$ is able to utilize $\mathcal{A}$ to solve the $q-SDH$ problem. The probability that $\mu^*\notin\{\mu_1, \mu_2, \cdots, \mu_{q-1}, \mu\}$ is $ (1-\frac{q}{p}) $. \\
 $\bm{ Case-II:\ \theta^*\notin\{\theta_1, \theta_2,.. \theta_q\}, \mu^*=\mu_i}$.\\
In this case, we have $ {K_2^*}= (g_0\cdot (g_2 \cdot B) ^{w^*}\cdot g_2^{\theta^*}) ^{\frac{1}{z+\mu^*}}, { (K_2) }_i= (g_0\cdot (g_2 \cdot B) ^{w_i}\cdot g_2^{\theta_i}) ^{\frac{1}{z+\mu_i}}$. Given $\mu^*=\mu_i, K_2^*={ (K_2) }_i$,  we obtain 
\begin{align*}
g_0\cdot (g_2 \cdot B) ^{w^*}\cdot g_2^{\theta^*}&=g_0\cdot (g_2 \cdot B) ^{w_i}\cdot g_2^{\theta_i}\\
 (g_2 \cdot B) ^{w^*}\cdot g_2^{\theta^*}&= (g_2 \cdot B) ^{w_i}\cdot g_2^{\theta_i}\\
 (g_2 \cdot B) ^{w^*-w_i}&=g_2^{\theta_i-\theta^*}\\
B&=g_2^{\frac{\theta_i-\theta^*-w^*+w_i}{w^*-w_i}}
\end{align*}
$\mathcal{B}$ can compute $\log_{g_2}B={\frac{\theta_i-\theta^*-w^*+w_i}{w^*-w_i}}$ by using $\mathcal{A}$. Therefore,  $\mathcal{B}$ is able to break the DL assumption by using $\mathcal{A}$. The probability of this case can be computed as $\frac{1}{p}\cdot \frac{1}{p} \cdot  (1-\frac{1}{p}) =\frac{p-1}{p^3}$\\
\item {\bf Type-II:} The user identity related to the key has been previously queried,  namely,  $\theta^*\in\{\theta_1, \theta_2,.. \theta_q\}$.  We take the two cases as follows into consideration.\\
$\bm{ Case-III:\ \theta^*\in\{\theta_1, \theta_2,.. \theta_q\}, \mu^*\notin\{\mu_1, \mu_2,.. \mu_{q-1}, \pi\}}$. \\
$\mathcal{B}$ can compute $ (\mu, g^{\frac{1}{z+\mu}}) $ in the same way as {\it Case-I} to break the $q-SDH$ assumption with the probability that $\mu^*\notin\{\mu_1, \mu_2,.. \mu_{q-1}, \mu\}$ is $ (1-\frac{q}{p}) $. \\
$\bm{ Case-IV:\ \theta^*\in\{\theta_1, \theta_2,.. \theta_q\},  \mu^*\in\{\mu_1, \mu_2,.. \mu_{q-1}, \pi\}, K_2^*={ (K_2) }_i}.$\\
The probability that $\mu^*=\mu$ is $\frac{1}{q}$. Because $\mu\notin\{\mu_1, \mu_2,.. \mu_{q-1}\}$,  $\mathcal{B}$ can compute $ (\mu, g^{\frac{1}{z+\mu}}) $ in the same way as {\it Case-I} to solve the $q-SDH$ problem. \\
For the sake of completeness of the proof,  we need to analyse the advantage $\mathcal{B}$ posses in solving the $q-SDH$ problem.  The probabilities of  {\it Case-I}, {\it Case-II}, {\it Case-III} and {\it Case-IV} forgeries are denoted as $\Pr[{\it Case-I}]$, $\Pr[{\it Case-II}]$,  $\Pr[{\it Case-III}]$ and $\Pr[{\it Case-IV}]$ respectively. The four cases are independently and identically distributed, each occurring with a probability of $\frac{1}{4}$.  Hence,  the advantage of breaking the $q-SDH$ assumption and $DL$ assumption by $\mathcal{B}$ can be calculated as follows. 
\begin{align*}
&\Pr[\text{Type-I}]\!=\! \Pr[{\it Case-I}]+\Pr[{\it Case-II}]\!=\!\frac{\epsilon_1}{4}\left(1-\frac{q}{p}\right)\!+\!\frac{\epsilon_2}{4}\frac{p-1}{p^3}\\
&\Pr[\text{Type-II}]\!=\!\Pr[{\it Case-III}]+ \Pr[{\it Case-IV}]\!=\!\frac{\epsilon_1}{4}\left(\left(1-\frac{q}{p}\right)\!+\!\frac{1}{q}\right)\\
&\epsilon= \max\left\{\Pr[\text{Type-I}],\Pr[\text{Type-II}]\right\}\\
&=\max\left\{\frac{\epsilon_1}{4}\left(1-\frac{q}{p}\right)+\frac{\epsilon_2}{4}\frac{p-1}{p^3}, \frac{\epsilon_1}{4}\left(\left(1-\frac{q}{p}\right)\!+\!\frac{1}{q}\right)\right\}
\end{align*}
\end{itemize}
\end{proof}
\subsection{Privacy Preservation}
{\theorem The PPKeyGen algorithm depicted in Figure 3 satisfies both ${\bf leakage-freeness}$ and {\bf selective-failure-blindness} under the $DL$ assumption. }\label{pp}With the intention of proving this theorem, two lemmas as follows are needed.\\
{\lemma Under the DL assumption, the PPKeyGen algorithm explicated in Figure 3 is selective-failure blindness. }\label{sfb}\\
\begin{proof}[Proof]KGC is malicious and tries to distinguish the user's identity $\theta$ embedded in the key. $g_0, g_1, g_2$ are generators in group $\mathbb{G}$, $\mathcal{BG}(1^\lambda ) \rightarrow (e, p, \mathbb{G}, \mathbb{G}_T)$. KGC selects $\vec{s}= (s_1, s_2,\cdots, s_l) \stackrel{R}{\leftarrow}\mathbb{Z}_p^l$ and computes $\vec{h}=\{h_i=g_1^{s_i}\}\ for\ i \in [l]$.Tracer selects $b \stackrel{R}{\leftarrow}\mathbb{Z}_p$ randomly, calculating $B=g_2^b$. KGC selects a random $a$ and publishes $Y=g_0^a$. KGC sets $msk=(a,s_{i})$ as master secret key, publishing $ (\vec{h}, Y) $. The tracer's secret-public key pair is $(b, B)$. Public parameters of the system can be denoted as $PP= (e, p, \mathbb{G}, \mathbb{G}_T, g_0, g_1, g_2, B, $ $Y, h_1, \cdots  ,h_l)$.   KGC submits $ (\theta_0, \vec{y}) $ and $ (\theta_1, \vec{y}) $ and selects $\delta\in\{0, 1\}$ randomly.  KGC can utilize the user oracles $U (PP, \theta_\delta, decom_\delta)$ and $U (PP, \theta_{1-\delta},$ $ decom_{1-\delta})$.
Subsequently,  KGC and an honest user U perform the protocol illustrated in Figure 3.  The oracle U will output $sk_0 (\theta_0, \vec{y}) $ and $sk_1 (\theta_1, \vec{y}) $. 
\begin{align*}
&1)  sk_\delta=\perp, sk_{1-\delta}=\perp,  U\ \text{returns}\  (\zeta, \zeta) \ \text{to}\ KGC;\\
&2)  sk_\delta=\perp, sk_{1-\delta}\neq\perp,  U\ \text{returns}\  (\perp, \zeta) \ \text{to}\ KGC;\\
&3)  sk_\delta\neq\perp, sk_{1-\delta}=\perp,  U\ \text{returns}\  (\zeta, \perp) \ \text{to}\ KGC;\\
&4)  sk_\delta\neq\perp, sk_{1-\delta}\neq\perp,  U\ \text{returns}\  (sk_0, sk_1) \ \text{to}\ KGC.
\end{align*}
In PPKeyGen,  the user computes $A_1, A_2$,  generates $\Sigma_U=PoK\{\left (w_1, \theta, \tau\right) :{A}_{1}=h^{\tau}\cdot {B}^{w_1}\wedge A_{2}={ (g_2\cdot B) }^{w_1}\cdot g_2^\theta\}$ and sends $A_1, A_2, \Sigma_U$ to KGC. Until this stage,  KGC executes either one or both oracles whose perspective on the two oracles remains computational indistinguishable,  because of the hiding property of commitment schemes and the zero-knowledge property of zero-knowledge proofs.  When KGC has the ability to compute the $k= (K_1, K_2, K_3, K_4, K_5) $ for the first oracle,  it can predict $k_\delta$ without using the steps below. 
\begin{itemize}
\item Firstly,  KGC verifies $\Sigma_K=PoK\{\left (a, { (s_i) }_{i \in [l]}\right)\!:$$B_1=g_0^{\langle\vec{y}, \vec{s}\rangle}\cdot$$A_1^{\frac{1}{d+a}} \cdot B^{\frac{w_2}{d+a}}\wedge B_2=g_0^{\frac{1}{d+a}}$$\cdot A_2^{\frac{1}{d+a}}\cdot  (g_2\cdot B) ^{\frac{w_2}{d+a}}\wedge B_3=g_1^{\frac{1}{d+a}}\wedge e (B_3, g_1^{B_5}\cdot Y) =e (g_1, g_1)  \wedge e (B_4, g_1^{B_5}\cdot Y) =e (g_1, h) \}$.
If it is invalid,  KGC returns $k_0=\perp$. 
\item For the second oracle, KGC outputs another $\chi= (B_1, B_2, B_3, B_4, B_5) $ and generates $\Sigma_K=PoK\{\left (a, { (s_i) }_{i \in [l]}\right) :B_1=g_0^{\langle\vec{y}, \vec{s}\rangle}\cdot A_1^{\frac{1}{d+a}} \cdot B^{\frac{w_2}{d+a}}\wedge B_2=g_0^{\frac{1}{d+a}}\cdot A_2^{\frac{1}{d+a}}\cdot  (g_2\cdot B) ^{\frac{w_2}{d+a}}\wedge B_3=g_1^{\frac{1}{d+a}}\wedge e (B_3, g_1^{B_5}\cdot Y) =e (g_1, g_1)  \wedge e (B_4, g_1^{B_5}\cdot Y) =e (g_1, h) \}$. If it is invalid,  KGC returns $k_1=\perp$. 
\item If both of the two steps above are successful,  then KGC proceeds as follows:
\begin{align*}
&1)  k_0=\perp, k_1=\perp,  U\ \text{returns}\  (\zeta, \zeta) \ \text{to}\ KGC;\\
&2)  k_0=\perp, k_1\neq\perp,  U\ \text{returns}\  (\perp, \zeta) \ \text{to}\ KGC;\\
&3)  k_0\neq\perp, k_1=\perp,  U\ \text{returns}\  (\zeta, \perp) \ \text{to}\ KGC.
\end{align*}
\item If $k_0\neq\perp, k_1\neq\perp$,  KGC returns $\theta_0$ and $\theta_1$.  KGC aborts when any of them fails,  otherwise it returns $ (k_0, k_1) $. 
\end{itemize} 
The prediction of $sk_0 (\theta_0, \vec{y}) $ and $sk_1 (\theta_1, \vec{y}) $ is right and consistent with
the oracle's distribution.  Consequently,  if both proofs are successful,  by implementing $PPKeyGen(KGC\!\leftrightarrow\!U)$, KGC can generate a valid secret key that U possesses. Therefore, if KGC can forecast the output of  $U (PP, \theta_\delta, decom_\delta) $ and $U (PP, \theta_{1-\delta}, decom_{1-\delta}) $, KGC's advantage in distinguishing between the two oracles is the same as the probability of no interaction. 
Therefore,  the advantage of KGC should come from the received $A_1, A_2$ and the proof $\Sigma_U=PoK\{\left (w_1, \theta, \tau\right) :{A}_{1}=h^{\tau}\cdot {B}^{w_1}\wedge A_{2}={ (g_2\cdot B) }^{w_1}\cdot g_2^\theta\}$.  Due to the witness indistinguishability property of zero-knowledge proofs and the hiding property of commitment schemes,  KGC is incapable of distinguishing between $U (PP, \theta_\delta, decom_\delta) $ and $U (PP, \theta_{1-\delta}, decom_{1-\delta})$ with a non-negligible advantage. 
\end{proof}
{\lemma The PPKeyGen algorithm,  as delineated in Figure 3,  is leakage-free. }\label{lf}
\begin{proof}[Proof]Assuming the existence of a malicious user U in the real-world experiment,  U interacts with an honest KGC executing the {\bf PPKeyGen} protocol. A corresponding simulator  $\mathcal{S}$ can be established in the ideal experimental,  which has access to the honest KGC performing  {\bf KeyGen} algorithm in ideal world. $\mathcal{S}$  conveys the input of $\mathcal{D}$ to U and U's output to $\mathcal{D}$, simulating the interaction between $\mathcal{D}$ and U.  Process of the real-world experiment is shown below. 
\begin{itemize}
\item The simulator $\mathcal{S}$ sends public parameters $PP$ to malicious user U. 
\item U computes $A= (A_1, A_2) $ and generates $\Sigma_U=PoK\{\left (w_1, \theta, \tau\right) :{A}_{1}=h^{\tau}\cdot {B}^{w_1}\wedge A_{2}={ (g_2\cdot B) }^{w_1}\cdot g_2^\theta\}$. $\mathcal{S}$ aborts if the proof fails.  Otherwise,  $\mathcal{S}$ can rewind $\left (w_1, \theta, \tau\right)$  from $\Sigma_{U}$. 
\item $\mathcal{S}$ sends $\theta$ to the honest KGC,  executes {\bf KeyGen} to generate $ (K_1, K_2, $\\$K_3, K_4, K_5) $. 
\item $\mathcal{S}$ computes $B_4=K_4, B_1=K_1\cdot{B_4^\tau}, B_2=K_2, B_3=K_3, B_5=K_5$.
\end{itemize}
We assume ${ (K_1, K_2, K_3, K_4, K_5) }$ is a valid key generated by the honest KGC under ideal-world experiment, while ${ (B_1, B_2, B_3, B_4, B_5) }$ is a valid key from KGC under real-world experiment; furthermore,  ${ (B_1, B_2, B_3, B_4, B_5) }$ distributes identically, and so is ${ (K_1, K_2, K_3, K_4, K_5) }$.  Therefore,  $\mathcal{D}$ cannot distinguish real-world experiment from real-world experiment. 
\end{proof}

\section{Comparison and Implementation}\label{sec5}

\subsection{Comparison and Efficiency Analysis}
\rm
We conduct a performance comparison, including communication cost and computation cost of our PPTFE-IP scheme with existing schemes~\cite{dpp:tfeip20,lawh:tfeip22,dpsm:feip22,lawy:feip24} in this subsection.  We mainly consider the computationally intensive operations such as exponential,  pairing and hash operations,  while disregarding other operations. Additionally, for notational simplicity, we define the following notations: $l$ represents the dimension of vectors; $|\mathbb{G}_1|$,  $|\mathbb{G}_T|$ and $|\mathbb{Z}_p|$ represent the length of element respectively on group $\mathbb{G}$, $\mathbb{G}_T$ and $\mathbb{Z}_p$;  $E_x$ and $E_{x_T}$ represent the cost of performing one exponential operation on $\mathbb{G}$ and $\mathbb{G}_T$;  $E_{H}$ represents the time of performing one hash operation,  and $P$ stands for the time of performing one pairing operation.

Table~\ref{comm1} presents the computational costs of the three schemes:~\cite{dpp:tfeip20}, {\it Trace-and-Revoke FE-IP scheme under DDH} described in Section 5.2 of~\cite{lawh:tfeip22} and ours. Table~\ref{comp1} shows communication costs of the three schemes.  From Table~\ref{comm1}, with the assumption that $l\gg1$, we know that the communication cost of our {\bf Setup} and {\bf Encryption} is lower than the other two schemes, while {\bf Key Generation} is slightly more expensive than~\cite{dpp:tfeip20,lawh:tfeip22}. Moreover, as observed in Table~\ref{comp1}, the computation cost of our traceable FE-IP scheme is about the same as~\cite{dpp:tfeip20,lawh:tfeip22}.
\begin{table}[h]
\scriptsize
\begin{threeparttable}
\caption{Communication Cost Comparison between Existing Work and Our TFE-IP Scheme in Figure 2}\label{comm1}
\begin{tabular*}{\textwidth}{@{\extracolsep\fill}lcccc}
 \toprule%
Scheme&Setup& Key Generation & Encryption \\
\midrule
~\cite{dpp:tfeip20}& $(2l+1)|\mathbb{Z}_p|+(l+1)|\mathbb{G}_1|+(l+1)|\mathbb{G}_T|$ & $|\mathbb{G}_2|$ & $l|\mathbb{G}_T|+l|\mathbb{G}_1|$ \\
~\cite{lawh:tfeip22}\tnote{1}& $ (l+2)|\mathbb{G}|+2l|\mathbb{Z}_p|$& $l|\mathbb{S}|+2|\mathbb{Z}_p|$ & $l|\mathbb{Z}_p|+ (l+2) |\mathbb{G}|+l|\mathbb{S}|$\\
\bf{Ours} &\bm{$ (l+5) |\mathbb{G}_1|+ (l+2) |\mathbb{Z}_p|$}&\bm{$2|\mathbb{Z}_p|+3|\mathbb{G}_1|$}&\bm{$ (l+3) |\mathbb{G}_1|+|\mathbb{Z}_p|$}\\
\bottomrule
\end{tabular*}
\begin{tablenotes}
\item[1]{In \cite{lawh:tfeip22},  $|\mathbb{S}|$  means the size of one element in the public directory pd, namely, a vector of size $l$.}
\end{tablenotes}
\end{threeparttable}
\end{table}
\begin{table}[h]
\scriptsize
\begin{threeparttable}
\caption{Computation Cost Comparison between Existing Work and Our TFE-IP Scheme in Figure 2}\label{comp1}
\begin{tabular*}{\textwidth}{@{\extracolsep\fill}lccccc}
\toprule%
Scheme&Setup&Encryption&KeyGen&Decryption&Trace \\
\midrule
~\cite{dpp:tfeip20}\tnote{1}& $lE_{x_1}+lE_{x_T}$ & $lE_{x_1}+2lE_{x_T}$ & $  E_{x_2}$ & $ lE_{x_1}+lE_{x_T}$&$\frac{8\lambda N^2}{\mu(\lambda)}\cdot$\\
&$+P$&&&$+P$&$\left(lE_{x_1}+2lE_{x_T}\right)$ \\
&&&&&\\
~\cite{lawh:tfeip22}\tnote{1}& $2lE_x$& $ (2l+2) E_x$ &0&$ (l+3) E_x$&$\frac{(N+1)\lambda N^2}{\mu(\lambda)}\cdot$\\
&&&&&$(2l+2) E_x$\\
&&&&&\\
\bf{Ours}\tnote{2}&\bm{$ (l+2) E_x$}&\bm{$ (2l+3) E_x$}&\bm{$9P+ (l+11) E_x$}&\bm{$ (l+2) E_x+5P$}&\bm{$4P+3E_x$}\\
\bottomrule
\end{tabular*}
\begin{tablenotes}
\item[1] In~\cite{dpp:tfeip20,lawh:tfeip22}, $N$ means user number, $\lambda$ represents security parameter and $\mu(\lambda)$ is a non-negligible function of $\lambda$ .
\item[2]This cost includes an extra cost of Key Verification,  while the other schemes don't include the Key Verification function. \\
\end{tablenotes}
\end{threeparttable}
\end{table}

\begin{table}[h]
\scriptsize
\begin{threeparttable}
\caption{Communication Cost of {\bf PPKeyGen} Algorithm in Figure 3}\label{comm2}
\begin{tabular*}{\textwidth}{@{\extracolsep\fill}lcc}
\toprule%
Scheme&User&KGC \\
\midrule
Ours\tnote{1}&$5|\mathbb{G}_1|+11|\mathbb{Z}_p|$&$ (4l+5) |\mathbb{Z}_p|+ (l+8) |\mathbb{G}_1|$\\
\bottomrule
\end{tabular*}
\begin{tablenotes}
\item[1]{This cost of {\bf PPKeyGen} Algorithm includes the extra communication cost of zero-knowledge proof.}
\end{tablenotes}
\end{threeparttable}
\end{table}

\begin{table}[h]
\scriptsize
\begin{threeparttable}
\caption{Computation Cost of {\bf PPKeyGen} Algorithm in Figure  3}\label{comp2}
\begin{tabular*}{\textwidth}{@{\extracolsep\fill}lcc}
\toprule%
Scheme&User&KGC \\
\midrule
Ours\tnote{1} &$ (3l+25) E_x+2E_H$&$ (l+18) E_x+2E_H$\\
\bottomrule
\end{tabular*}
\begin{tablenotes}
\item[1]{This cost includes the extra computation cost of zero-knowledge proof. Because multiplying several discrete logarithms with the same base can be viewed as adding their exponents together, i.e.$\prod\limits_{i=1}^l  (g_0^{y_i}) ^{s_i}=g_0^{\sum\limits_{i=1}^l{y_i\cdot s_i}}$, so the cost of $\prod \limits_{i=1}^l  (g_0^{y_i}) ^{s_i}$ and $\prod_0^l{\left ( ({g_0}^{\mu_{a_i}}) \cdot  ({g_0}^{d\cdot y_i}) \right) ^{s_i^\prime}}$ calculated by KGC can be considered as one exponential operation. 
In $\Sigma_K$,  $\prod_0^l{\left ( ({g_0}^{\mu_{a_i}}) \cdot  ({g_0}^{d\cdot y_i}) \right) ^{\tilde{s_i}}}$ computed by the user costs $(1+l)E_x$ because the user doesn't know $\mu_{a_i}$.}
\end{tablenotes}
\end{threeparttable}
\end{table}
\subsection{Implementation and Evaluation}

Our TFE-IP scheme introduced in Figure 2 and the PPKeyGen algorithm depicted in Figure 3 are implemented and evaluated.  The proposed PPTFE-IP scheme is realized in Ubuntu20. 04 (64 bit) system on an Intel (R)  Core (TM) i7-8550U CPU with 8G of RAM. The PBC library~\cite{pbc} in Linux is utilized in order to perform bilinear pairing operations.  And we use socket programming for two-party secure computation.  This program is implemented in C language on Linux.                                   

When implementing our scheme,  we consider the following five cases: $l=10, l=20, l=30, l=40$ and $l=50$,  and each time is obtained by taking the average value after 10 experiments.  Figure  4 depicts the time spent by the algorithms in each stage of our PPTFE-IP scheme. 

As shown in Figure 4, the {\bf Setup} algorithm takes 0.0341988s,  0.0547218s,  0.0695288s,  0.0871566s and 0.1064724s in each case.  The  {\bf Encryption} process costs 0.0429196s, 0.0748678s, 0.1075483s, 0.1420852s and 0.1828628s for each case. {\bf Encryption} takes more time than {\bf Setup},  but slightly less than twice {\bf Setup} time. 

\begin{figure}[h]%
\centering
\includegraphics[width=0.9\textwidth]{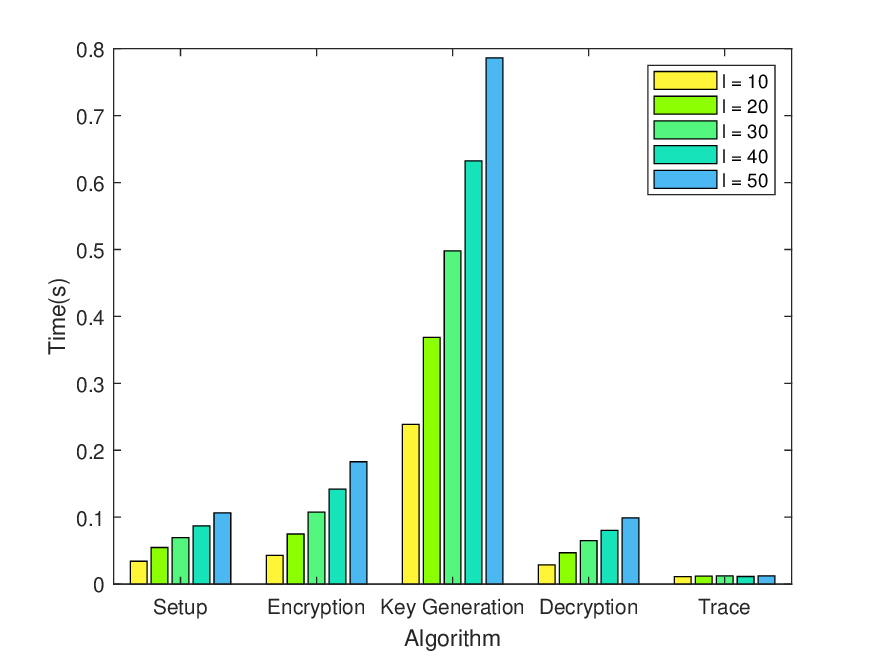}
\caption*{Figure 4: The Computation Cost of Our PPTFE-IP Scheme}\label{fig4}
\end{figure}

Moreover, Figure  4 shows that the {\bf Key Generation} algorithm takes a longer execution time than the other algorithms. Our experiment shows that the {\bf Key Generation} algorithm  (including key verification)  takes 0.2387588s,  0.3687528s,  0.4979975s,  0.6324328s,  0.7862466s time in the five cases, respectively.  Due to the two party secure computing, the {\bf PPKeyGen} algorithm takes more time.  On the other hand,  multiple pairing and exponentiation operations are required during the key verification,  which are time-consuming. 

Especially, in each cases the {\bf Decryption} algorithm costs 0.0286769s,  0.0468638s,  0.0649096s,  0.0802697s and 0.098845s, which is faster than {\bf Setup}.  We use the socket to simulate the communication between the KGC and a user.

The computation cost of the {\bf Setup} algorithm,  {\bf Encryption} algorithm, {\bf Key Generation} algorithm and {\bf Decryption} algorithm all grows linearly with the dimension of vectors. Conversely, the {\bf Tracing} algorithm takes 0.0113066s,  0.0119908s,  0.0122832s,  0.0114053s,  0.0124318s in the five cases, respectively. It is indicated that our {\bf Tracing}  algorithm is efficient since it is independent of the vector dimension.

\section{Conclusion and Future Work}\label{sec6}


To protect users' privacy and realize traceability, we introduced the first privacy-preserving traceable functional encryption (PPTFE-IP) scheme. Specifically, we presented the concrete construction of TFE-IP and the {\bf PPKeyGen} algorithm. Additionally, a detailed security proof of PPTFE-IP was offered.  Furthermore,  we conducted a comparative analysis of our scheme with existing traceable FE-IP schemes, implementing and evaluating the efficiency of our scheme.

However, the{\bf PPKeyGen} algorithm in our scheme is computationally expensive. Therefore, constructing a PPTFE-IP scheme with efficient key generation algorithm is interesting and challenging. This will be our future work.





\bibliographystyle{elsarticle-num} 
    \bibliography{sn-bibliography}

\appendix
\section{The Details of Zero-Knowledge Proof}\label{app}

\subsection{The Detail of $\Sigma_U$}\label{A_U}
An instantiation of the proof $\Sigma_U$ is given as follows. 
\begin{itemize}
\item User selects\ $\tau^\prime, \theta^\prime, w_1^\prime \stackrel{R}{\leftarrow}\mathbb{Z}_p$, and computes
\begin{align*}
&{A}_{1}^\prime  =h^{\tau^\prime}\cdot {B}^{w_1^\prime},  \nonumber \\
&A_2^\prime = (g_2\cdot B) ^{w_1^\prime}\cdot g_2^{\theta^\prime}, \\
&c=H_1\left (A_1||A_1^\prime||A_2||A_2^\prime\right) , \nonumber \\
&\tilde{\tau}=\tau^\prime-c\cdot \tau, \nonumber \\
&\tilde{\theta}=\theta^\prime-c\cdot \theta\nonumber,  \\
&\tilde{w_1}=w_1^\prime-c\cdot w_1, 
\end{align*}
\item User sends $\left (A_1,A_1^\prime,A_2,A_2^\prime\right) $ and $\left (c, \tilde{\tau}, \tilde{\theta}, \tilde{w_1}\right) $to verifier (KGC). 
\item After receiving $\left (A_1,A_1^\prime,A_2,A_2^\prime\right) $ and $\left (c, \tilde{\tau}, \tilde{\theta}, \tilde{w_1}\right) $, KGC verifies
\begin{align*} 
&c\overset{\text{?}}{=}H_1\left (A_1||A_1^\prime||A_2||A_2^\prime\right) , \nonumber \\
&{A}_{1}^\prime \overset{\text{?}}{=}h^{\tilde{\tau}}\cdot {B}^{\tilde{w_1}}\cdot A_1^c,  \nonumber \\
&{A}_{2}^\prime \overset{\text{?}}{=} (g_2\cdot B) ^{\tilde{w_1}}\cdot g_2^{\tilde{\theta}}\cdot A_2^c. \\
\end{align*}
\end{itemize}

\subsection{The Detail of $\Sigma_K$}\label{A_K}
An instantiation of the proof $\Sigma_K$ is given as follows. 
\begin{itemize}
\item KGC selects\ $ (s_i^\prime) _{i\in[l]}, a^\prime \stackrel{R}{\leftarrow}\mathbb{Z}_p$,  and computes
\begin{align*}
& ({g_0}^{\mu_{a_i}}) ^\prime  = (g_0^{y_i}) ^{ (a^\prime) }, {\mu_{a_i}}={y_i}\cdot a\\
& (\frac{B_1^d}{A_1\cdot B^{w_2}}) ^\prime=B_1^{-a^\prime}\cdot \prod_0^l\left ( ({g_0}^{\mu_{a_i}}) \cdot  ({g_0}^{d\cdot y_i}) \right) ^{ (s_i^\prime) }, \\
& (\frac{g_0\cdot A_2 \cdot  (g_2\cdot B) ^{w_2}}{B_2^d}) ^\prime=B_2^{ (a^\prime) }, \\
& (\frac{g_1}{B_3^d}) ^\prime=B_3^{ (a^\prime) }, \\
& (\frac{h}{B_4^d}) ^\prime=B_4^{ (a^\prime) }, \\
&c=H_1\left ( ({g_0}^{\mu_{a_i}}) ^\prime||B_1|| (\frac{B_1^d}{A_1\cdot B^{w_2}}) ^\prime||B_2|| (\frac{g_0\cdot A_2 \cdot  (g_2\cdot B) ^{w_2}}{B_2^d}) ^\prime||B_3|| (\frac{g_1}{B_3^d}) ^\prime||B_4|| (\frac{h}{B_4^d}) ^\prime||B_5\right) , \nonumber \\
&\tilde{a}=a^\prime-c\cdot a,  \\
&\tilde{s_i}={s_i}^\prime-c\cdot{s_i}\ for\ i\in[l],  \\
\end{align*}
\item KGC sends $\left ( ({g_0}^{\mu_{a_i}}) ^\prime,B_1,(\frac{B_1^d}{A_1\cdot B^{w_2}}) ^\prime,B_2, (\frac{g_0\cdot A_2 \cdot  (g_2\cdot B) ^{w_2}}{B_2^d}) ^\prime,B_3, (\frac{g_1}{B_3^d}) ^\prime,B_4,B_5\right) $ and $\left (c, \tilde{a}, \tilde{ (s_i) }_{ i\in[l]}\right) $to verifier (User). 
\item After receiving $\left ( ({g_0}^{\mu_{a_i}}) ^\prime,B_1, (\frac{B_1^d}{A_1\cdot B^{w_2}}) ^\prime,B_2, (\frac{g_0\cdot A_2 \cdot  (g_2\cdot B) ^{w_2}}{B_2^d}) ^\prime,B_3, (\frac{g_1}{B_3^d}) ^\prime,B_4,B_5\right) $ and $\left (c, \tilde{a},  ({\tilde{s_i}}) _{i\in[l]}\right) $,  User verifies
\begin{align*}
&c\overset{\text{?}}{=}H_1\left ( ({g_0}^{\mu_{a_i}}) ^\prime||B_1|| (\frac{B_1^d}{A_1\cdot B^{w_2}}) ^\prime||B_2|| (\frac{g_0\cdot A_2 \cdot  (g_2\cdot B) ^{w_2}\cdot}{B_2^d}) ^\prime||B_3|| (\frac{g_1}{B_3^d}) ^\prime||B_4||B_5\right) , \nonumber \\
& ({g_0}^{\mu_{a_i}}) ^\prime \overset{\text{?}}{=} (g_0^{y_i}) ^{\tilde{a}}\cdot  ({g_0}^{\mu_{a_i}}) ^c, \\
& (\frac{B_1^d}{A_1\cdot B^{w_2}}) ^\prime \overset{\text{?}}{=}B_1^{-\tilde{a}}\cdot \Pi_0^l \left ( ({g_0}^{\mu_{a_i}}) \cdot  ({g_0}^{d\cdot y_i}) \right) ^{\tilde{s_i}}\cdot  (\frac{B_1^d}{A_1\cdot B^{w_2}}) ^c, \\
& (\frac{g_0\cdot A_2 \cdot  (g_2\cdot B) ^{w_2}}{B_2^d}) ^\prime\overset{\text{?}}{=}B_2^{\tilde{a}}\cdot  (\frac{g_0\cdot A_2 \cdot  (g_2\cdot B) ^{w_2}\cdot g_2^\theta}{B_2^d}) ^c, \\
& (\frac{g_1}{B_3^d}) ^\prime\overset{\text{?}}{=}B_3^{\tilde{a}}\cdot  (\frac{g_1}{B_3^d}) ^c, \\
& (\frac{h}{B_4^d}) ^\prime\overset{\text{?}}{=}B_4^{\tilde{a}}\cdot  (\frac{h}{B_4^d}) ^c. \\
\end{align*}
\end{itemize}
\end{document}